\documentclass{article}

\usepackage{multirow}

\usepackage{arxiv}

\usepackage[utf8]{inputenc} 
\usepackage[T1]{fontenc}    
\usepackage{hyperref}       
\usepackage{url}            
\usepackage{booktabs}       
\usepackage{amsfonts}       
\usepackage{nicefrac}       
\usepackage{microtype}      
\usepackage{lipsum}		
\usepackage{graphicx}
\usepackage{natbib}
\usepackage{doi}
\usepackage[affil-it]{authblk}

\title{Value Prediction for Spatiotemporal Gait Data using Deep Learning}

\author[1]{Ryan Cavanagh}
\author[1, *]{Jelena Trajkovic}
\author[1]{Wenlu Zhang}
\author[2]{I-Hung Khoo}
\author[3]{Vennila Krishnan}

\affil[1]{%
  Department of Computer Engineering and Computer Science,  	California State University Long Beach}
\affil[2]{%
  Electrical Engineering and Biomedical Engineering,  	California State University Long Beach}
\affil[3]{%
    Department of Physical Therapy,
    California State University Long Beach}
\affil[*]{Corresponding author: Jelena.Trajkovic@csulb.edu}

\hypersetup{
pdftitle={Cavanagh,Trajkovic et al. Gait Value Prediction},
pdfsubject={Cavanagh,Trajkovic et al. Gait Value Prediction},
pdfauthor={Cavanagh,Trajkovic et al.},
pdfkeywords={Gait Value Prediction; Recurrent Neural Network; Convolutional Neural Network; Time-series Gait Data; Health Monitoring; Rehabilitation; Wearable Technology},
}

\begin{document}
\maketitle

\begin{abstract}
Human gait has been commonly used for the diagnosis and evaluation of medical conditions and for monitoring the progress during treatment and rehabilitation. The use of wearable sensors that capture pressure or motion has yielded techniques that analyze the gait data to aid recovery, identify activity performed, or identify individuals. Deep learning, usually employing classification, has been successfully utilized in a variety of applications such as computer vision, biomedical imaging analysis, and natural language processing. We expand the application of deep learning to value prediction of time- series of spatiotemporal gait data. Moreover, we explore several deep learning architectures (Recurrent Neural Networks (RNN) and RNN combined with Convolutional Neural Networks (CNN)) to make short- and long-distance predictions using two different experimental setups. Our results show that short-distance prediction has an RMSE as low as 0.060675, and long-distance prediction RMSE as low as 0.106365. Additionally, the results show that the proposed deep learning models are capable of predicting the entire trial when trained and validated using the trials from the same participant. The proposed, customized models, used with value prediction open possibilities for additional applications, such as fall prediction, in-home progress monitoring, aiding of exoskeleton movement, and authentication.

\end{abstract}


\section{Introduction}
\label{sec:intro}
Clinical gait analysis has been used in the diagnostic, assessment, monitoring, and prediction of health conditions \cite{Baker2016}. For example, human gait can be used as an important diagnostic indicator to discern pathological conditions such as stroke, Parkinson's disorders, and other neuro-musculoskeletal conditions, from normal. Traditionally, to establish a diagnosis of the gait disorder, a trained clinician would either analyze a participant using observational gait analysis, quantitatively record (pathological) gait via a motion capture system, or record spatiotemporal gait parameters. While trained clinicians provide needed expertise, their observation, and analysis are subjective and they are usually not available outside of the clinical environment or for an extended time (e.g. for in-home rehabilitation or monitoring). Therefore, quantitative gait measurement and analysis are valuable aids in diagnosis as well as in the prognosis of a disorder as they can facilitate objective monitoring and data acquisition for long time periods. 

Recent innovations in technology allow quantitative gait data acquisition, that removes subjectivity and the need for prolonged clinical observation by an expert. Apart from the gold-standard motion capture system, which is expensive and labor-intensive~\cite{MobilityLab}, a variety of sensor-based systems have been used to monitor or analyze gait. For example, devices that sense pressure and motion are attached to feet (Walk-Even insole \cite{Khoo_2015}, SmartShoe \cite{Bae_2009, Kong_2008}) or knees, thighs, and hips \cite{Mannini_2011} to record the spatiotemporal parameters of gait. Innovation in technology also expands applications beyond clinical use. For example, gait analysis has been used for biometric identification \cite{Potluri_2019, Moon_2020}, for detecting impaired driving \cite{Li2019}, sports activity \cite{Ghazali2018CommonSA}, or (general) human activity  \cite{Lee_2019}. Hence, contemporary sensor-based monitoring and data acquisition systems facilitate a more objective, quantitative assessment of gait, while being portable, accurate, easily available, and less expensive.

The above advances in gait data acquisition have enabled new applications based on statistical and machine-learning models (see Section \ref{sec:related}). However, there still remains a large gap in gait modeling and analysis that is tuned to the individual under observation, is easy to extend to new individuals, and can be used in real-time applications. We address this gap by proposing a technique to generate \textit{accurate} and \textit{customized} deep-learning models to predict spatiotemporal gait data for \textit{each} individual. Our models can be used to aid professionals in rehabilitation or prognosis, in clinical and in-home environments. Target applications include qualitative and quantitative feedback to post-stroke survivors \cite{Lee2019Stroke}, aiding exoskeleton motion \cite{Schutter_Exoskeleton} or fall prediction \cite{Schniepp_fall_prediction_2021}. Moreover, as our models are customized to a particular individual, they can be used for authentication by providing a \textit{quantitative metric} of (dis)similarity of spatiotemporal gait parameters helping us identify a participant as one of known (in the dataset) or unknown individuals (not in the dataset).

The specific contributions of this work are:
\begin{enumerate}
    \item [C.1] We generate a set of deep-learning model architectures for short-distance and long-distance \textit{value prediction} for sensor data corresponding to human gait. While other works use gait-related data for classification, we can successfully predict actual values. Our models that target long-distance prediction can predict the \textit{entire trial}. Not only the proposed models \textit{expand beyond classifiers}, but they are also suitable to provide quantitative data for use in patient recovery, fall risk assessment, fall prediction, and in aiding exoskeleton movement.     
    \item [C.2] We propose architectures that can train \textit{individually} on the data from either each trial or from multiple trials from each participant, thus generating the models with the following advantages:
    \begin{enumerate}
        \item [C.2.a] Unlike traditional deep-learning models, which are trained on data with a large number of features, all our models are trained on data with a smaller number of features. Despite this challenge, the proposed approach yields \textit{customized models} that have been shown to have \textit{high accuracy}.     
        
        \item [C.2.b] We show that our models can predict the values in a \textit{fraction of the time} needed for a participant to make a move and generate new data. Therefore, the proposed models can be used in \textit{real-time applications} that require high accuracy, such as patient monitoring. 

        \item  [C.2.c] In contrast to the majority of other applications, there is \textit{no need to re-train one large unified model} using all training data when a new participant is added to the study. We only need to train a single new model for the new participant, thus saving on training time. 
    \end{enumerate}
   
    \item [C.3] We \textit{expand the application} of deep learning models (ensemble CNN+RNN models) to predict \textit{time-series spatiotemporal gait data values}. This is an application beyond typical use for the analysis of image caption data, and financial data.
\end{enumerate}

\subsection{Overview of the Proposed Method}
\label{sec:methodology}

We utilize a dataset obtained from Walk-Even insoles~\cite{Khoo_2015}, where each shoe insole is equipped with six sensors. Therefore, each trial records six sensor values from each foot as the participant applies pressure while walking at a regular pace. We predict the consecutive pressure values that correspond to each of the six sensors on the right foot. 

We propose a technique to train four deep learning model architectures for each individual to predict the pressure values of gait cycles: RNN with a single RNN layer, LSTM (Long Short Term Memory)  with two LSTM layers, Bidirectional LSTM with two bidirectional LSTM layers, and an ensemble CNN+RNN combining two 1-Dimensional convolutional layers and an LSTM layer. The first three models are designed to predict short sequential data, while the CNN+RNN model is designed to predict long sequential data. Usually, CNN+RNN architectures are designed to deal with image captioning analysis but in this project, we leverage CNN+RNN architecture to preserve temporal features in long-distance gait data. Specifically, CNN is used to shorten the input data while maintaining the overall patterns. Working with temporal data sets, combining CNN and RNN has been shown to improve prediction outcomes while lowering computational costs compared to traditional RNN models.

We present two experimental setups, `by trial' and `by participant'. Setup `by trial' is applied to a single gait trial to explore the ability of different models to make a short-distance prediction. Setup `by participant', is applied to multiple concatenated trials, from a single participant, to explore the ability of different CNN+RNN models to make a long-distance prediction, i.e. the ability to predict entire trials.
We evaluate model accuracy using standard evaluation metrics, MAE (Mean Absolute Error),  MSE (Mean Squared Error), and RMSE (Root Mean Squared Error), to calculate how closely the predicted time series data matches the actual time series (sensor) data. Both setups train on data obtained from the six sensors for the right foot that, in turn, predict the values for the same sensors in the next time steps. MAE, MSE, and RMSE values are computed using predicted values for each sensor. To provide a visual comparison and showcase the quality of prediction,  the sensor data for individual sensors (i.e. true test data values)  are summed up at each time step and plotted alongside the sum of predicted data for the corresponding time step. Our results show that the best-performing model for a short sequence prediction is Bidirectional LSTM with an average RMSE = 0.124346 and a minimum RMSE = 0.060675. The best-performing model for long-distance prediction in the vast majority of cases is CNN+RNN with an average RMSE = 0.194171, and as low as RMSE = 0.106365. In a few cases, for an output window of 3 or 4 Bidirectional LSTM slightly outperforms CNN+RNN, but at the expense of increased complexity and run-time. Moreover, the proposed models for long-distance prediction can accurately predict an entire trial for a participant, when trained and validated using the trials from the same participant. 

The rest of this paper is structured as follows: Section \ref{sec:related} reviews the previous works, and Section \ref{sec:data_acquisition} provides the description of the dataset. Section \ref{sec:techique} describes proposed modeling, and Section \ref{sec:experiments} presents the experimental results. We follow up with a discussion of limitations and future work (Section \ref{sec:limitations_and_future}), and conclude in Section \ref{sec:conclusion}.

\section{Related Work}
\label{sec:related}
Gait is the result of a cyclic series of movements and is characterized by the fundamental unit `gait cycle'. A gait cycle is defined as the sequence of events starting when a  foot contacts the ground to when that same foot contacts the ground again. Typically, during human locomotion, the right foot is taken as the reference limb. A full gait cycle could be divided into two major phases: stance and swing. The stance phase occurs as the reference foot is on the ground, supporting the body's weight. The swing phase occurs as the foot is in the air, being advanced for the next contact with the ground. At normal walking speed by a healthy participant, the stance phase occupies approximately 60\% of the gait cycle, and the swing phase occupies the remaining 40\%.

There have been numerous studies on gait, gait cycle analysis and classification,  the applications of sensors (pressure, Inertial Measurement Unit (IMU), or other wearable sensors), and motion capture systems, to gait. We present the below studies that relate to the proposed work and point to possible future directions.

\subsection{Gait Cycle Analysis, Deterministic and Statistical Modeling}
\label{sec:gait_cycle}
The initial efforts that utilized sensor data for gait analysis used deterministic and statistical methods to analyze or classify the gait cycle or its phases. For example, Tong et al. \cite{Tong_1999} analyzed data from a gyroscope attached to the ankle to distinguish between straight and curved walking using correlation coefficients. Sabatini et al. \cite{Sabatini_2005} analyzed angular change reported by an IMU sensor, to distinguish between gait cycle phases. 

Moreover, the knowledge obtained from sensors collecting gait information can be applied to evaluate a person's gait~\cite{Koldenhoven_2018-journal,Murai_2018, Khoo_2017}. Koldenhoven et al.~\cite{Koldenhoven_2018-journal} conducted a validation study to compare gait cycle data obtained using wearable sensors (triaxial accelerometer and gyroscope) and the 3D motion capture system. They showed that the runner's data could accurately be captured using wearable sensors outside traditional laboratory settings. Murai et al.~\cite{Murai_2018} use IMU and a large-scale system (the commercial marker-based optical motion capture system with 15 cameras) to build a software model, using correlation coefficients. The model is then used with IMUs only, outside the lab, to assess risks for runner's injury. Both systems show that wearable devices, with appropriate software, can provide reliable metrics for human locomotion. 

Khoo et al. \cite{Khoo_2017} also used gait cycle data to re-train post-stroke individuals using in-sole pressure sensors with real-time biofeedback. They measured the weight distribution of an individual's foot and determined the asymmetry of gait. They used a deterministic algorithm for the gait phase analysis. We utilize the same data obtained by \cite{Khoo_2017} but instead of deterministic algorithms for re-training we use deep learning modeling for value prediction for human gait. 

Bae et al.~\cite{BAE_journal_2011} utilized data collected from embedded air-bladder force sensors in a shoe insole to analyze the gait phases of healthy individuals. Their method builds a Hidden Markov Model (HMM) using the sensor-obtained values to detect gait phases and check for abnormal state transitions between the phases. Mannini et al. \cite{Mannini_2011} developed an HMM applicable to data collected from a foot-mounted gyroscope. The model detects the gait phases for individuals walking at different speeds and inclinations on the walking surface. Such systems enable accurate and automatic detection of gait cycles, which are essential for thorough gait analysis in normal as well as pathological populations. Our study differs concerning the objective of making predictions and the prediction method used. Both aforementioned studies use a statistical model, HMM, for gait phase detection, while the proposed method uses deep learning on time series of gait data values for value prediction. 

\subsection{Machine Learning-Based Gait Analysis and Detection Algorithms }
\label{sec:ml_gait}

The use of sensors to better understand human gait has been merged with machine learning \cite{Costilla-Reyes2020, Wang2019, Potluri_2019, Moon_2020}. Costilla-Rayes et al. \cite{Costilla-Reyes2020} present applications of deep learning in security and healthcare. Wang et al. \cite{Wang2019} survey the recent advance of deep learning-based sensor-based activity recognition. Parkka et al. \cite{Parkka_2006} obtained data from various sensors (e.g. gyroscope, accelerometers, and physiological sensors) attached to eight body parts with the participants at different settings (e.g. walking or climbing stairs). They formulated machine learning models (decision tree and neural network) to classify one of seven activities that a participant is performing. Lee et al. \cite{Lee_2019} used the data from pressure and IMU sensor readings obtained from FootLogger insole~\cite{Footlogger} to construct a separate Convolutional Neural Network model for each type of input (sensor) to perform classification. The dataset in both studies has a significantly larger number of features than in our study. A large number of features usually makes for easier training of accurate machine and deep learning models. We design the models for using fewer features that can achieve high prediction accuracy. Unlike previous work which focuses on classification, we predict actual values in time-series data, setting us apart. 

Moon et al. \cite{Moon_sensors_2020} showed that gait information can be used to distinguish between different individuals. FootLogger insoles~\cite{Footlogger} were used to record data from pressure and IMU sensors as the participants walked. They proposed a deep learning architecture that combined convolutional (CNN) and recurrent neural networks (RNN) to identify any of the tested participants. While the general idea of CNN+RNN architecture is similar to ours, the applications greatly differ. Moon et al.  \cite{Moon_sensors_2020} aim to identify a person based on their previous gait data, while we can predict an entire trial based on previous gait data. Similarly to the previous studies, we work with much-reduced input data size, making the prediction modeling additionally challenging, yet successful.

Some studies found that customized (subject-specific) models have higher accuracy of prediction than unified (generic) models. Zhang et al.\cite{Zhang2020} discovered that subject-specific models outperformed the genetic ones even though they did not need additional subject-specific characteristics. Similar findings have been reported for CNN models for a freeze of gait (FoG) detection \cite{xia2018evaluation}. The intuition behind those results is that gait spatiotemporal data are unique for each individual (please see Figures~\ref{fig:trial_truncated} and~\ref{fig:trial-2}), and furthermore, the data patterns might be different for the same individual between different/repeated trials (please see Figures~\ref{fig:trial-2} and~\ref{fig:trial-3}). This motivated us to propose a technique that generates customized models. Furthermore, having customized models removes the need to retrain the global model. Retraining requires the use of the entire training data set, as in \cite{Chen2015ADL, Moon_sensors_2020} which consumes both processing time and memory, and it is likely not suitable for real-time applications. Contrary, the proposed, customized models would need to train only a single trial (for 'by trial' setup) or on all available trials for the newly added participant (for 'by participant' setup). Finally, the run time of model generation (training) and prediction (testing) are more efficient with the proposed approach due to the use of customized models.

Horst et al. \cite{Horst2019ExplainingTU} devised a general framework that facilitates the understanding and interpretation of non-linear machine learning methods in gait analysis. They feed portions of the model's predictions back to the input variables to distinguish the variables that are most relevant for the characterization of gait patterns from a certain individual. The applications of sensor data analysis are not only limited to gait analysis but can also be applied to other areas of the body. Rawashdeh et al. \cite{Rawashdeh_2016}  used an IMU sensor attached to the upper arm of a person to track the motion of their shoulder to help prevent overuse injuries. They were able to classify the different positions of the shoulder and assess if repeated usage of the same form can cause injuries soon. Although not used in our methods, the characterization of inputs that are most influential to the output and the use of sensors positioned at different body parts are suitable directions for extensions of our work.

\subsection{Value Prediction for Human Motion using Machine Learning and Deep Learning}
\label{sec:value_prediction}
Rudenko et al. \cite{Rudenko2020} surveyed and proposed a taxonomy for human motion trajectory prediction. While human motion can have many notions, such as "full-body motion, gestures/facial expressions, moving through space by walking, using a mobility device or driving a vehicle," only "moving through space by walking" may relate to our study, as it includes a prediction of the values of human motion, such as trajectory coordinates or velocity profile. Human motion trajectory prediction has started to utilize Machine Learning and Deep Learning for value prediction. For example, Saleh et al. \cite{Saleh2018} used LSTM architecture to predict the trajectories of pedestrians. The prediction models were trained and tested on (sequential) images from the moving vehicle's stereo cameras. Djuric et al. \cite{Djuric2018} proposed a CNN-based uncertainty-aware vehicle motion prediction approach that utilizes high-definition map images containing the projected prior motion of the vehicle and surroundings. The model was capable of producing a short-distance trajectory of the vehicle alongside uncertainty. Finally, ensemble models that combine CNN and RNN (namely LSTM) have been used in this domain to combine "the spatial and temporal relations of the observed agent's motion" \cite{Rudenko2020}. Xue et al. \cite{Xue2018LSTM} introduced a hierarchical model where CNN was used to extract relevant information from the person's trajectory, social neighborhood, and global scene layout, and provide it to the LSTM model. Zhao et al. \cite{zhao2019multi} apply similar architecture on fused images of the environment and agents (both pedestrians and vehicles), thus capturing the interactions between the agents and forming a prediction. While previous deep learning architectures are similar to the proposed one, the nature of the input data, thus the application of those architectures, is very different. The motion trajectory prediction mainly uses sequences of images, thus having much more input (training) data and features. We carefully designed our models, through experimentation, to have high accuracy despite the small number of features. 

Ogata and Matsumoto \cite{Ogata2019} proposed a method to estimate the joint angles of the upper human body. They collected data using a wearable suit implanted with strain sensors. They utilized a CNN model to estimate the joint angle values for bending and straightening an elbow. Utilization of similar sensing mechanisms, and corresponding prediction modeling, is a promising direction for our future work, namely the application of value prediction for aiding exoskeleton movement. 

\subsection{Dataset Size}
\label{sec:dataset_size}
The number of participants also referred to as a sample size, their age, and health status are challenges for many machine learning and deep learning studies related to gait. Recent surveys \cite{Harris2022,Saboor2020LatestRT,Lingmei2019} illustrate the challenge. 
Harris et al. \cite{Harris2022} point out that a simple size is a concern across surveyed work: "With the clinical nature of these studies and the impaired gait participants they require, the barriers to experimentally collecting sufficient data are understandable". As per Lingmei et al. \cite{Lingmei2019}, the issue is even more pronounced for fall detection and prevention since falls are very rare events. Harris et al., also, cite potential solutions to obtaining data that represents a larger number of participants and a wider range of health conditions. From the usage of medications to induce fatigued gait in healthy subjects \cite{Lasselin_2020}, to the use of generated, synthetic data \cite{Arifoglu_2017} that reflect
features similar to abnormal. Finally,  the recent trend is to take
data acquisition out of the lab \cite{Russell_2021}. Cheng et al.  \cite{Cheng2017HumanAR} used passive monitoring, using a smartphone over 24 weeks, which is very time-consuming both for data acquisition and processing. Additionally, the proposed workaround strategies, such as  "aggregating, pre-processing, and
learning the data", address cost, time, 
user privacy, and clinical constraints, hence, the resulting systems may be somewhat simple and 
introduce errors \cite{Harris2022}. 

Saboor et al. \cite{Saboor2020LatestRT} recognize the issue and illustrate its magnitude. Out of 33 surveyed papers from 2015 to 2020, 59\% have a sample size of fewer than 30 participants (34\% have 1 - 10, and 25\% have 11 - 30). Only about 30\% have more than 100 participants, and roughly a third of those collect the data using a device that is placed in the pocket or attached to a hand, indicating the use of an IMU sensor or a smartphone. While the use of an IMU sensor or a smartphone allows for a larger number of participants, longer monitoring, and ease of setup outside of the clinical environment, the accuracy of this approach has been reported as a challenge. Therefore, we propose to extend our approach to the use of IMU sensors and strain sensors in addition to the pressure sensors to expand applications but strive for accuracy. Complementary to the issue of dataset size is the issue of how the data is acquired while using sensor-based systems. The data acquisition and therefore accuracy of models derived from the data will depend on sensor location and their installation on the body \cite{McGinnis2017, Zhao2017}, hence we will research sensor type and location needed for the proposed systems to be accurate.

\section{Dataset}
\label{sec:data_acquisition}

The data has been gathered in a controlled laboratory environment using a custom-made wearable device, called Walk-Even~\cite{Khoo_2017}. Walk-Even insoles are equipped with Force Resistive Sensors (FSR) placed on the front and back of the foot.  Each insole has a total of six FSR pressure sensors: three on the heel (‘back insole’) and three on the front part of the foot (‘front insole’). The three sensors are placed on top/bottom of each insole, but to the center and each side (toward the outer or inner edge of the foot). Walk-Even stores the raw data, produced by each sensor,  in a control unit, worn on the waist of the participant. The data has been gathered using healthy human participants. The data was obtained from 17 participants, with different physical characteristics, and gender, and with an age range from 18 to 28. The Institutional Review Board at the California State Univerity Long Beach approved the study protocol\footnote{IRB protocol number 639747-12} which conformed to the principles of the Declaration of Helsinki. All participants gave written informed consent before participation.

During the experiment (data acquisition), participants walked for 5 meters while wearing Walk-Even insoles on both feet.  Data is recorded for each of the 12 sensors (6 on each foot) in pound-force (lbf) every 8ms (i.e. with a sampling rate of 125 Hz). Each data point is given a unique ID and a time step value. Hence, the 12 sensors generate a time series of pressure values. Each participant was recorded for 3 to 12 trials, totaling 108 valid trials. Moreover, the trials have different total lengths, in terms of time and therefore a different corresponding number of samples, as each participant walked the same distance but may have had different speeds and stride lengths.

\begin{figure}[th]
    \centering
    \includegraphics[width=1.0\linewidth]{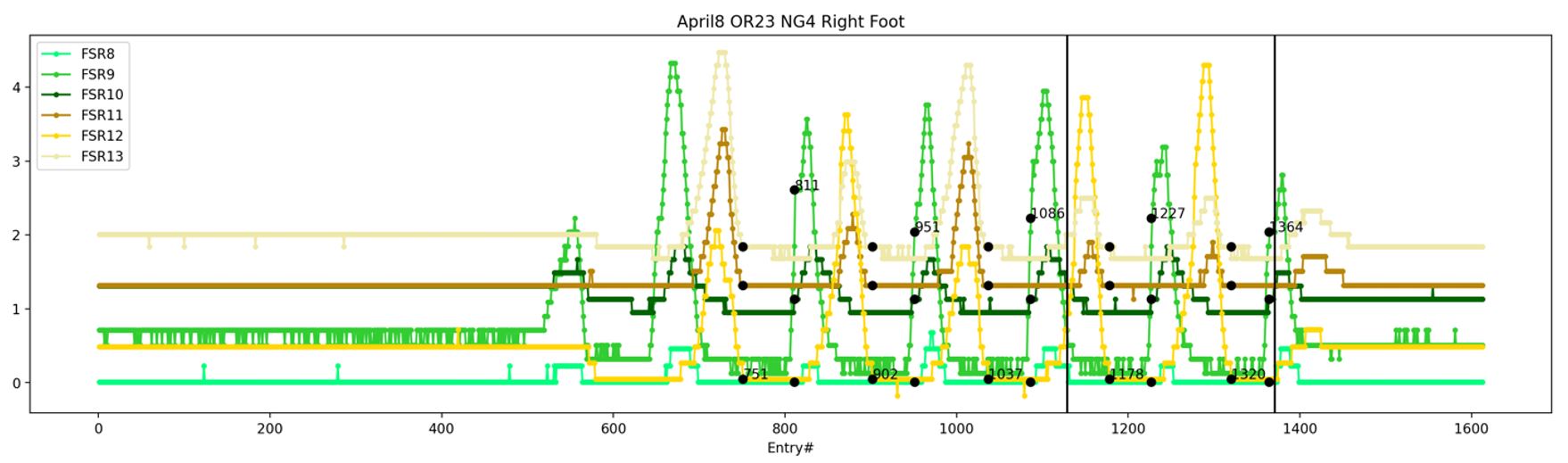}
    \caption{Example of a Trial: Plot of raw data values for each of six sensors on the right foot for participant  NG.}
    \label{fig:trial_raw}
\end{figure}

Figure \ref{fig:trial_raw} shows a plot of raw data from one trial for participant NG. The plot shows the pressure value recorded for each sensor on the right foot. High values reported by force-sensitive resistive sensors (FSR) correspond to high pressure on the corresponding sensor when (a part of) the foot is on the ground, while low values correspond to the foot being in the air. Varying values of pressures are due to participant walking, thus shifting pressure from the back of the foot to the front, and due to placement of the sensors, being in the front, outer or inner edge, and back.
In this study, we use pressure values for the right foot (six sensors) as inputs to the proposed deep learning models, during training and validation, as well as true data values (ground truth) during testing. We propose two distinct experimental setups, ‘by trial’ and ‘by participant’ that evaluate the model’s ability to make short-distance and long-distance predictions, respectively, the details of which are presented in~\ref{sec:experimental_setup}.

\section{Proposed Technique}
\label{sec:techique}

\subsection{Data Preprocessing}
\label{sec:prepossessing}

\begin{figure}[th]
    \centering
    \includegraphics[width=1.0\linewidth]{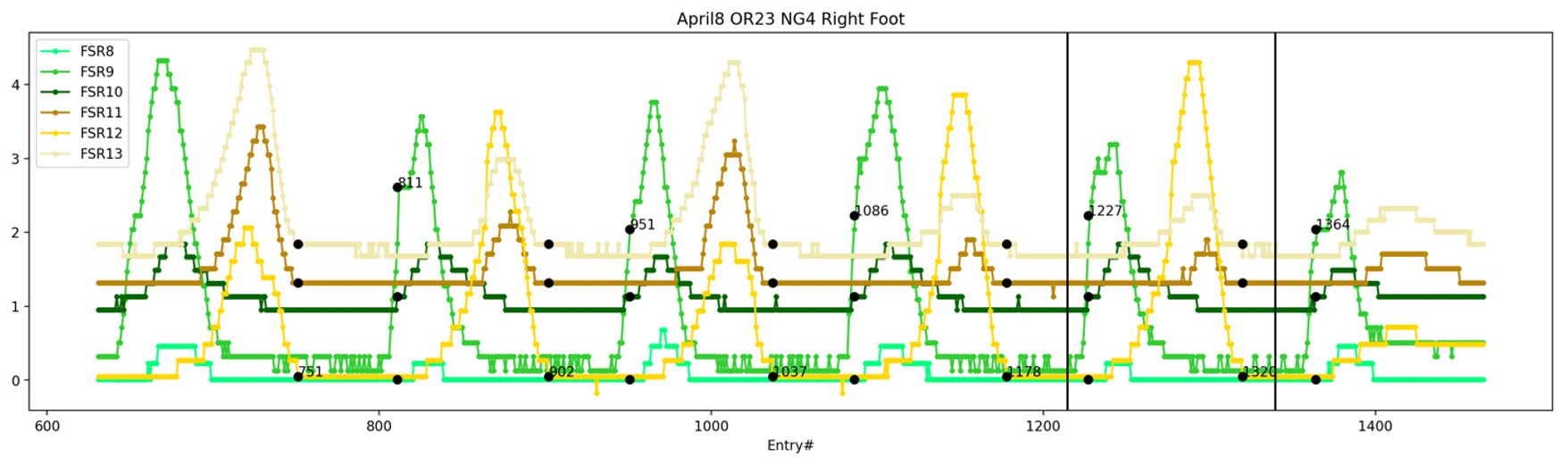}
    \caption{Example of a Trial: Plot of data values after preprocessing for each of six sensors on the right foot for participant  NG.}
    \label{fig:trial_truncated}
\end{figure}

As seen in Figure \ref{fig:trial_raw}, the data on the far left-hand side (time steps less than 600) and on the far right-hand side (time steps greater than 1500) of the plot, do not match the swing or stance phase. This data corresponds to sensor readings taken before and after the participant was instructed to walk, thus, this data should be removed. The preprocessing was done manually by careful visual observation of the plotted sensor values and their corresponding numerical values. We identified the time steps where the fluctuations in sensor data values start (for the beginning) and stop (for the end) to isolate the data values produced by walking. The resulting, truncated data is shown in Figure \ref{fig:trial_truncated}. The prepossessing ensures that data used to train, validate, and test the models represent true variations in sensor values that correspond to human gait. 

Please note that the data supplied for training, validation, and testing has been windowed. For example, an input window of size 5 has values for 5 time steps for each of the sensors. Each input window is used to train and validate the model to predict one or more output values. The window size is selected for each experimental setup (please see \ref{sec:experimental_setup}). Additionally, the data is normalized based on the training data for that specific experiment (for the `by participant' setup), meaning that an offset is added so that the minimum values are the same as for the training data. This is done for each trial to offset any differences between the trials and it serves instead of calibration.

\subsection{Modeling}
\label{sec:modeling}

Recently, deep learning has been a significantly popular area among the machine learning and artificial intelligence community due to its improved learning capabilities \cite{Janiesch2021}. Thanks to convolutional neural networks (CNN)~\cite{lecun1998gradient,krizhevsky2012imagenet,szegedy2015going,simonyan2014very} and recurrent neural networks (RNN)~\cite{kolen2001field, pascanu2013difficulty,pascanu2012understanding, karpathy2015visualizing}, deep learning architectures are implemented in various applications such as full self-driving system~\cite{Ni2020}, object detection~\cite{ren2015faster,he2017mask}, machine translation~\cite{vaswani2017attention}, and biomedical imaging analysis~\cite{zhang2021deep,zhang2015deep}. In this paper, we expand the application of several commonly used deep learning models to time-series gait data. We investigate the ability and quantify the accuracy of the customized models to predict different individuals' walking patterns. Namely, we explore their capability to predict pressure values created by the participant at every time step. We implement and evaluate four different models of short- and long-distance prediction. The first three models are a variation of the RNN model, hence in the analysis we may refer to them as "RNN models" whereas the last one is an ensemble CNN and RNN model. The model description is given below.

\textit{Recurrent Neural Networks (RNN)}~\cite{kolen2001field} models are intensively implemented to recognize feature patterns in a sequential dataset. Most RNNs are used in time-series data, language-related data, or biological DNA sequence data. One key attribute of RNNs is the ability to work with varying lengths of input. RNNs were designed to capture the hidden state features by using recurrent computation at each time step. The RNN cell sums the dot product of the input and the weights with the dot product of the previous output and the corresponding weights then passes the data through an activation function. Since the parameters are shared during the whole training time, simple RNNs have suffered from vanishing gradient issues\footnote{Vanishing gradient issues arise in models with gradient-based learning methods, where the weights for the model are computed interactively using the partial derivative of the error function \cite{Basodi2020_gradient}. When the gradient becomes vanishingly small, it will prevent the weight from changing from iteration to iteration, or even completely stop the training of the neural network.} especially if there is a long sequence of data, hence, the resulting model's accuracy suffers.

\textit{Long Short Term Memory (LSTM)}~\cite{Hochreiter1997long} model refers to the RNN model with LSTM layers.  The addition of LSTM cells helps alleviate the vanishing gradient problem. More precisely, the LSTM cells retain the input data, the previous cell's hidden state, and the cell state. Consequently, the LSTM model has greatly increased computational complexity but generates more accurate predictions.

\textit{Bidirectional LSTM (biLSTM)}~\cite{Schuster1997bidirectional} model refers to an improvement of the LSTM model where the input has been considered in a forward direction and in a backward direction. Hence, the data is passed through two LSTM layers separately: one where the original input is utilized (forward direction) and another where the input is reversed (backward direction). The weights from the two LSTMs are concatenated (or combined using other specified functions), allowing the model to predict with a knowledge of what has happened prior and what will happen in the future. Therefore, this model, also, has further increased complexity and further improved accuracy.

\textit{The (ensemble) CNN+RNN} model combines Convolutional Neural Network (CNN) and Recurrent Neural Network (RNN) models. CNN~\cite{krizhevsky2012imagenet,he2016deep} models are usually implemented for image classification problems due to the nature of convolutional layers. The convolutional layer, also known as the kernel or filter,  performs a dot product of the input and the weights to help identify patterns in the input data (usually patterns in the images). Designing different CNN architectures by manipulating convolutional, pooling, and fully connected layers and varying respective parameters will allow specific features or patterns to develop from the input data. This is how the proposed CNN architecture is customized for this problem. CNN models has also been used on time-series data, but for successfully predicting stock market movement ~\cite{Liu2022} and forecasting stock prices \cite{Tsantekidis2017}. Tsantekidis et al.~\cite{Tsantekidis2017} applied the CNN models, and Liu et al.~\cite{Liu2022} used CNN and RNN for time-series data (separately, not as an ensemble model), motivating us to use it in the proposed ensemble model for time-series gait data values. 

The intuition of combining CNN and RNN is that this model will fully address the vanishing gradient problem found in an RNN working with long sequences. Usually, CNN+RNN architectures are designed to deal with image captioning analysis but in this project, we leverage CNN+RNN architecture to preserve temporal features in long-distance gait data. Specifically, CNN is used to shorten the input data while maintaining the overall patterns, hence tackling the vanishing gradient problem. Working with temporal data sets, combining CNN and RNN has been shown to improve prediction outcomes while lowering computational costs compared to traditional RNN models. The CNN encodes the time-series features~\cite{lecun1998gradient}, then RNN decodes them to generate the next time-stamp sequences~\cite{shi2021neural}. The features extracted using CNN make for an excellent means to predict sequences of data such as translations, text generation, and time-series data, using RNN. 

A concern when using a simple RNN for long sequences of data is the tendency for the first inputs used for training to have minimal to no effect on the outputs. Instead of using a simple RNN, one solution to this problem is to use a long short-term memory (LSTM) layer. The proposed method, CNN+RNN, consists of two 1-dimensional CNN layers and an RNN layer with LSTM cells to solve the information loss of long-distance sequential data.

\begin{figure}[th]
    \centering
    \includegraphics[width=0.8\linewidth]{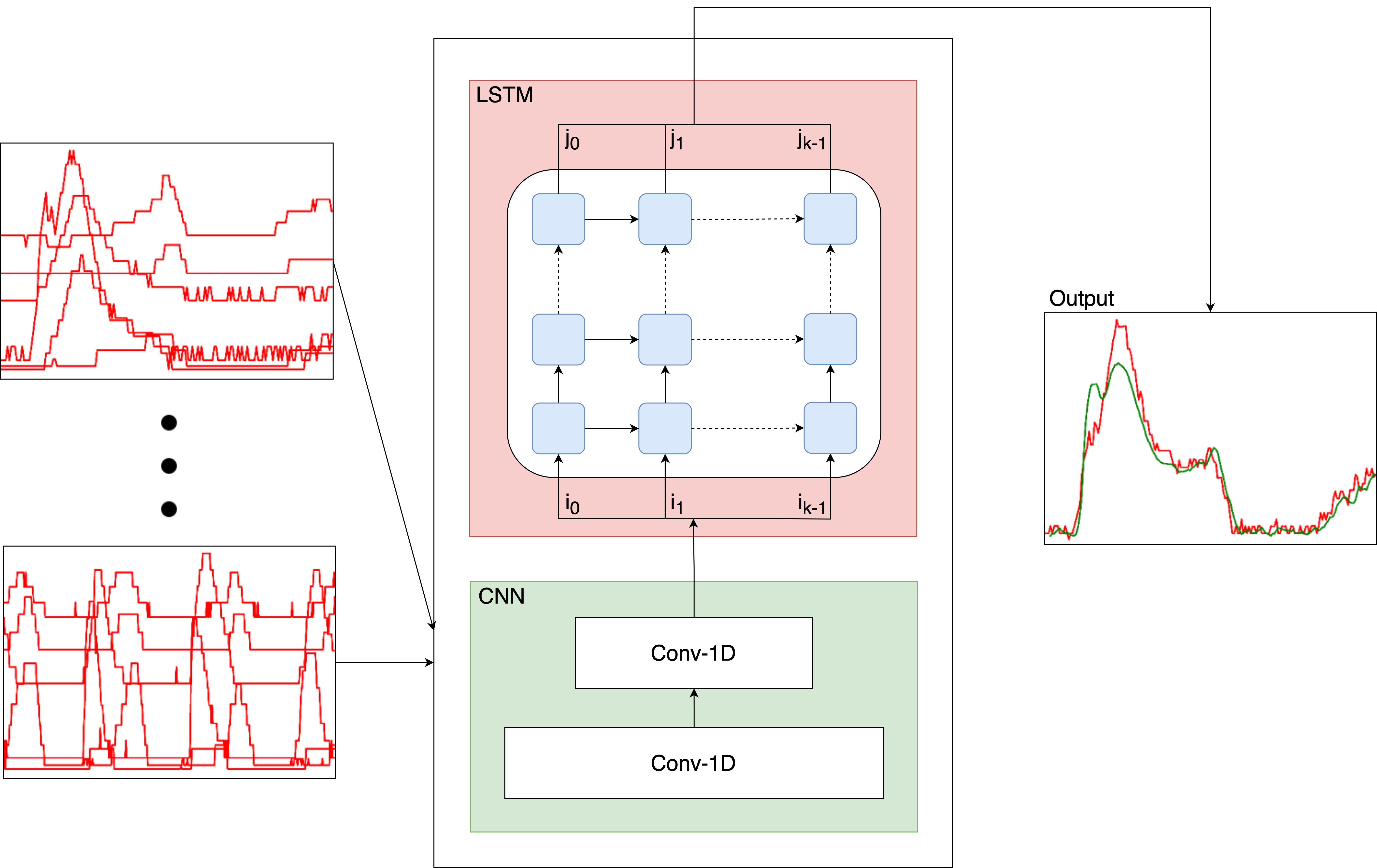}
    \caption{The proposed deep learning architectures, CNN+RNN, with two convolutional 1D layers (both with 20 filters each and kernel size 3), and one LSTM layer (with 20 units).} 
    \label{fig:CNN_RNN}
\end{figure}

Figure~\ref{fig:CNN_RNN} shows a block diagram of the proposed ensemble CNN+RNN model architecture when used for the setup `by participant'. On the left are examples of trials used as inputs: each plot shows time-series data for six sensor values for the duration of each trial. Even though the trials are shown separately, for the setup `by participant' the trials are concatenated and windowed as they are provided as input. On the right is an example of the output: a plot of the sum of all six predicted values (shown in green) for one entire trial plotted alongside the sum of all six values of sensors' reading, i.e. the true test data values (shown in red) for the same trial. The middle block represents the proposed custom model's architecture that has been designed and fine-tuned on our dataset by experimentation. First, the input passes through the CNN section, composed of two 1-dimensional convolutional layers both of which have 20 filters with the kernel size set to 3. The output from this section is passed to the RNN, which is an LSTM layer with 20 units, followed by a dense layer\footnote{The dense layer is not shown in Figure~\ref{fig:CNN_RNN}.}. Both convolutional layers in our proposed architecture have a kernel size of 3, which decreases the input, for example from 20 to 16. The LSTM layer takes these 16 data values as inputs and passes the outputs to the dense layer returning the prediction. 

The proposed architecture is capable of predicting the entire trial. A predicted value is generated for each sensor (for each time step defined by the output window size). The evaluation metrics are computed for each data point, thus comparing the predicted values to true test data values for each sensor.

The model is also used in the `by trial' setup. In this case, only a single trial is used as input. It is partitioned into the training and validation portions (for training), and a testing portion (to be compared to the predicted output, for testing).

\subsection{Evaluation Metrics}
\label{sec:eval_metrics}

To evaluate the performance of the several different deep learning models, we employ the commonly used metrics: Mean Absolute Error \big(MAE\big),  Mean Squared Error \big(MSE\big), and Root Mean Squared Error \big(RMSE\big). The equations use $y_i$ for the true test data value (i.e. the ground truth), $\hat{y}_i$ for the predicted value, and $N$ for the number of data values in the true test dataset.

MAE computes the average of the difference between the true test data values and predicted values through the dataset.

\begin{equation}
MAE = \frac{1}{N} \sum_{i=1}^{N} |\hat{y}_i - y_i|
\label{mae}
\end{equation}

MSE calculates the average of the squared difference between the true test data values and predicted values through the dataset.
\begin{equation}
MSE = \frac{1}{N}\sum_{i=1}^{N}(y_{i}-\hat{y}_i)^{2}
\label{mse}
\end{equation}

RMSE is the square root of MSE.
\begin{equation}
RMSE =\sqrt{MSE} =\sqrt{\frac{1}{N}\sum_{i=1}^{N}(y_{i}-\hat{y}_i)^{2}}
\label{rmse}
\end{equation}
MAE measures the average absolute distance from the prediction to the true test data values, MSE measures the vertical distance from the prediction to true test data values, and RMSE measures the variation between the predicted and true test data values. Therefore, we adopt \textit{all three} measurements to evaluate the performance of our prediction models. A model with a \textit{good fit} will have values \textit{close to zero} for all three metrics. A model with the perfect fit will have them all equal to 0.

\section{Experiments}
\label{sec:experiments}

\subsection{Experimental Setup}
\label{sec:experimental_setup}
We implement four deep learning models to predict sensor data values that correspond to an individual's gait cycle: a simple RNN with a single RNN layer, an LSTM model with two LSTM layers, a Bidirectional LSTM (biLSTM) with two bidirectional LSTM layers, and the CNN+RNN model combining two 1-dimensional convolutional layers and an LSTM layer, as described in Section~\ref{sec:modeling}. The first three models are built on top of a simple RNN model, which we may refer to as RNN models in our analysis, and they are designed to predict short-distance sequential data. The last model (CNN+RNN) is designed to explore the ability to predict long-distance sequential data. 

The proposed models are implemented in Keras 2.5.0 using the backend of Tensorflow 2.5.0 on the Google Cloud Platform with a two-core Intel Xeon CPU 2.30GHz. The optimizer for all RNN models is Adam \cite{Adam} with a learning rate set to 0.0001, whereas the optimizer for the CNN+RNN model is RMSprop \cite{RMSprop} with a learning rate set to 0.01. 
We also experimented with the use of dropout layers for all models. For all models, the first RNN layer used a recurrent dropout set to 0.5 and the CNN+RNN model used an additional dropout set to 0.1 for the LSTM layer. The use of a dropout layer produces results with inferior performance and hence we will not present them in the results section.

Typically, deep learning models are generated using a much larger number of features with very large training data sets. We face the challenge of having a smaller training set, containing 17 participants with 108 trials containing time-series sensor data for each of the six sensors, where each sensor value has been considered a feature. To overcome this challenge we carefully fine-tune model architectures and propose two different experimental setups: `by trial' and `by participant'. Additionally, for every experiment, the data is normalized based on the training data for that specific experiment; this is especially pertinent for the setup `by participant' due to the use of multiple trials for training, validation, and testing. Both experimental setups use data values for the right foot (6 sensor values at each time step) as input data and predict 6 values corresponding to each sensor, at \textit{one or more} consecutive (future) time steps. 

The input data is being windowed, and each input window produces predictions for as many time steps as defined by the output size. The input and output window sizes are specified for each model, in two experimental setups. The input window size defines the number of time steps whose sensor values are used as inputs for generating the prediction. We vary the input size in two experimental setups. The output window size defines the number of time steps for which the prediction is generated. Each model contains a dense layer that makes \textbf{6 x W} predictions, where \textit{6} corresponds to each predicted sensor value, and \textit{W} corresponds to the output window size. Two presented experimental setups have different values for the output window size. Since the `by trial' experimental setup explores short-distance prediction, the output window size is 1, hence, the prediction will be made for one time step for each sensor value (a total of 6 predicted values). Since the `by participant' experimental setup explores long-distance prediction, the output size is larger than 1. For example, when the output size is 3, the prediction will be made for three consecutive time steps, for each sensor value (a total of 18 predicted values). The predicted values for all sensors at each time step are summed, and graphed to easily compare the predicted values to (summed) true test data values. However, MAE, MSE, and RMSE values are computed using the predicted non-summed values and the true test data values.

\subsubsection{Setup by Trial - Experimental Configuration}
\label{sec:by_trial}
To explore the ability of all four models to perform short-distance prediction of sequential data this setup uses a single gait trial, split into approximately 70\% for training, 15\% for validation, and 15\% for testing data. To ensure that testing and validation data would be 15\% each, a function was created that allocated 15\% of the input data plus the remainder final window size of data to the testing data, consequently making the training set equal to or set slightly below 70\%.  Figure \ref{fig:trial_truncated} shows a truncated trial for participant NG. We show two black vertical lines towards the right side of the image (one to the right of time step 1200 and the other to the left of time step 1400) dividing the data into three segments. The largest segment to the left of the first black line corresponds to training data, the next segment corresponds to the validation, and the rightmost segment corresponds to the testing data set. If we were to compare the segments in Figure \ref{fig:trial_truncated} to the original data shown in Figure \ref{fig:trial_raw} we see that the testing data would have contained mainly constant values, thus giving an unfair advantage to the evaluation of the model. This is one of the reasons to preprocess the data as in Section \ref{sec:prepossessing}.

\begin{table}
  \caption{Time Parameters in Seconds for 'by trial' Setup: Average Training (and Validation) Time (Avg. Train), Average Prediction Time (Avg. Predict), Time to Generate True Test Data Values (Avg. Measured Test Time)}
  \label{tab:trial_runtimes}
  \centering
  \begin{tabular}{lccc}
    \toprule
    Model &Avg. Train Time [s] &Avg. Predict Time [s] & Avg. Measured Test Time [s]\\
    \midrule
    Simple   & 84.157   & 0.084& \multirow{4}{*}{0.922}\\
    LSTM     & 175.310  & 0.084&\\
    biLSTM   & 250.783  & 0.087&\\
    CNN+RNN  & 99.513   & 0.081&\\
  \bottomrule
\end{tabular}
\end{table}

Since this setup targets short-distance prediction, the input window size is 5 and the output window size is 1. Training setup `by trial' uses MSE as the loss function and trains for 40 epochs. All models for setup `by trial' are capable of predicting the data in under one-tenth of value generation time (using true test data values). Table \ref{tab:trial_runtimes} shows average values of time parameters for our experiments `by trial', reported in seconds.  The average training (and validation) time for all models spans between 84 and 251 sec, showing that the customized models take a very short time to train (below 5 min for the most complex one). As expected, Simple RNN has the shortest, while biLSTM has the longest training time. Our custom CNN+RNN architecture trains, on average, for 99.5 sec, which is just slightly longer compared to the training time of the simple RNN. Average prediction times are between 0.081 sec (CNN+RNN) and 0.084 sec (Simple RNN and LSTM). Similar to training, biLSTM takes the most time to predict (0.087 sec). Additionally, the average time a participant takes to walk the distance that corresponds to the testing portion of the trial (thus generating true test data values) is 0.922 sec. We see that all models are \textit{able to predict the values in real-time}, and the prediction time is more than ten (10) times shorter than the data generation/acquisition. 

\subsubsection{Setup by Participant - Experimental Configuration}
\label{sec:by_participant}
To explore the ability of all four models to perform long-distance prediction of sequential data, we introduce a second experimental setup that evaluates the different input and output window sizes for each participant, for all proposed deep learning models. For this experimental setup, one trial was allocated for testing, another for validation, and the remaining trials were first windowed and then concatenated together to form the training data set. The window size has an input size of 5, 10, 15, and 20 and an output size of 3, 4, and 5 for each input size. For each participant and for each model we explore a total of 12 input/output combinations (4 input and 3 output values). Training setup `by participant' uses MAE as the loss function and trains for 20 epochs.

All models in this setup are capable of predicting all true test data values for the entire trial within 0.12 seconds. Table \ref{tab:participant_runtimes} shows average values of time parameters for our experiments `by participant', reported in seconds.  The average training (and validation) time for all models spans between 62 and 265 sec, showing that the customized models, even for the long-distance prediction take a very short time to train, which is comparable to the time taken to train `by trial' models. As expected, Simple RNN has the shortest, while biLSTM has the longest training time. Our custom CNN+RNN architecture trains, on average, for 94.4 sec, which is about 50\% longer compared to the training time of the simple RNN. Average prediction times are between 0.082 sec (Simple RNN) and 0.111 sec (biLSTM). As expected, biLSTM takes the most time to predict (0.111 sec) and CNN+RNN takes close to the minimum time (0.088 sec). Additionally, the average time a participant takes to walk the distance that corresponds to test trial (thus generating true test data values) is 57.870 sec. We see that all models are \textit{able to predict the values in real-time}, and the prediction time is a small fraction of the time of the data generation/acquisition. 

\begin{table}
  \caption{Time Parameters in Seconds for 'by participant' Setup: Average Training (and Validation) Time (Avg. Train), Average Prediction Time (Avg. Predict), Time to Generate True Test Data Values (Avg. Measured Test Time)}
  \label{tab:participant_runtimes}
  \centering
  \begin{tabular}{lccc}
    \toprule
    Model &Avg. Train Time [s] &Avg. Predict Time [s] &Avg. Measured Test Time [s]\\
    \midrule
    Simple   & 61.302   & 0.082& \multirow{4}{*}{57.870}\\
    LSTM     & 153.448  & 0.101&\\
    biLSTM   & 265.008  & 0.111&\\
    CNN+RNN  & 94.352   & 0.088&\\
  \bottomrule
\end{tabular}
\end{table}

Please note that the `by participant' setup predicts the entire trial and, within it, values for each sensor at \textit{multiple} time steps, whereas the `by trial' setup predicts only 15\% of a trial and within it a single value for each sensor at \textit{one} time step. Therefore, CNN+RNN is shown to be faster than other models while predicting more accurately a larger amount of values than other models. 

\subsubsection{Participant Selection}
\label{sec:participant_selection}
Our current gait data are produced with healthy participants and we run experiments for both setups for all trials from all participants. This means that the `by trial' setup uses all 108 trials separately for a total of 108 experiments. Additionally, the `by participant' setup uses 108 trials that are combined into 17 groups, one for each participant. Moreover, it runs experiments for each input and output window combination for a total of 204 experiments.  Due to space limitations, we present aggregated data for each setup and only a subset of experiments showcasing the trials and participants that illustrate experiments that are closest to presented averages, the best-performing and the poorest-performing models. We also present trials and participants that give us some insight into the used model and experimental setup, or showcase lessons learned for the future. Hence, the trials and participants that we present were selected solely to show the results that illustrate our findings for each setup.

\begin{table}[th]   
    \centering
    \includegraphics[width=0.99\linewidth]{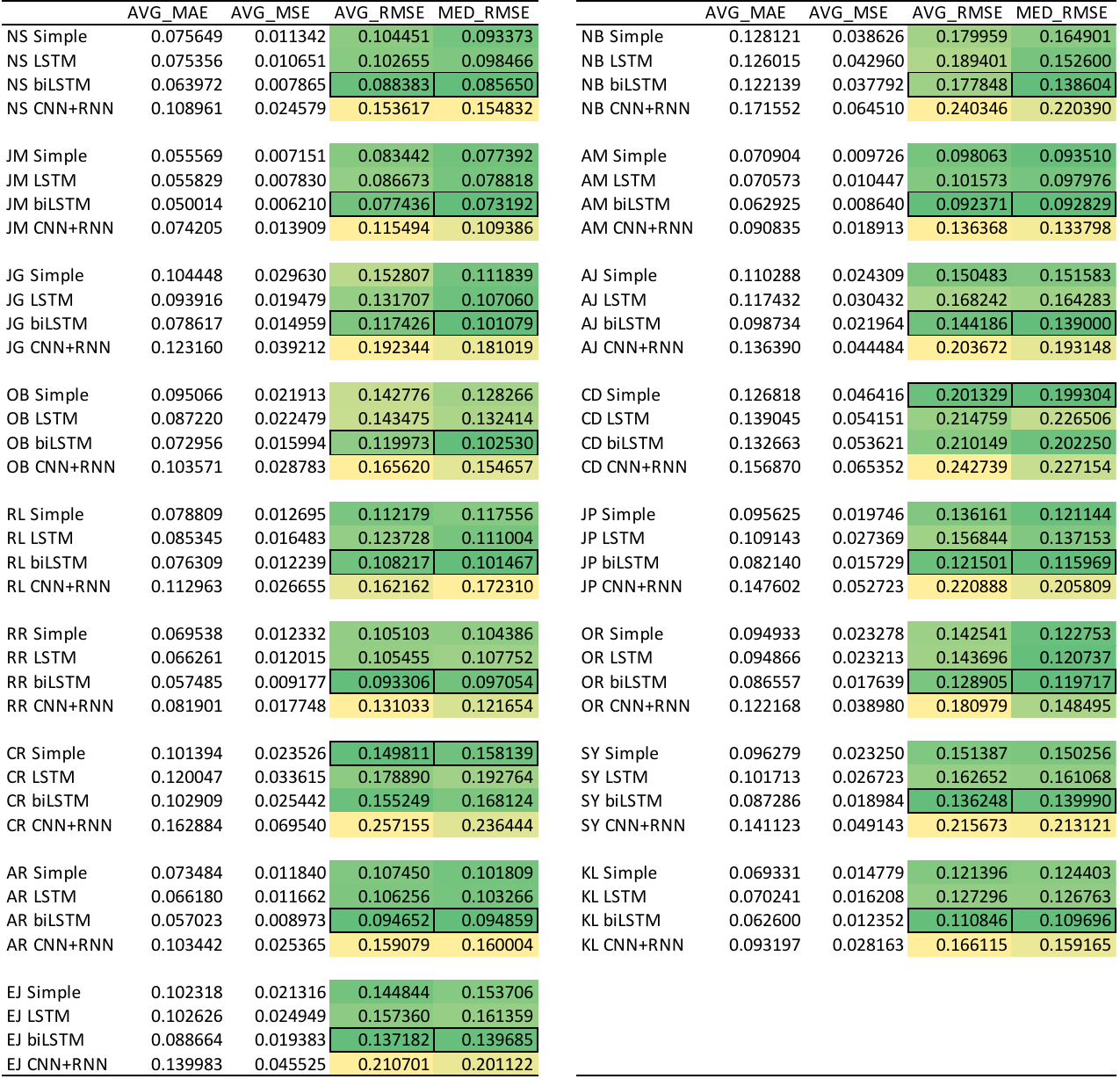}
    \caption{Setup by Trial: Average Error Metrics by Model}
    \label{tab:trial-avg-table}
\end{table}

\subsection{Results, Analysis, and Discussion}
\label{sec:results}
We present and compare the experimental results and visualization for predicting force-resistive sensor data values for the setups `by trial' and `by participant'. In the tables, we display the values of each evaluation metric for each model. We indicate the performance of the model by shading from dark green (best) to pale yellow (poorest). The graphs show the summed sensor values, and, as needed, individual sensor values for the true test data values and the predicted data values. In both types of graphs, the true test data values (i.e. the sensor readings) are shown in blue and the predicted values are shown in pink. The graphs that plot (raw) sensor data, which show separate values for each force-sensitive resistive sensor (FSR 8 -13), are added to aid the analysis of training and prediction. The type of graph presented for each experiment was selected to facilitate an understanding of the model's ability or obstacles while making accurate predictions. 

\subsubsection{Setup by Trial}
\label{sec:exp_by_trial}

The average scores for all three metrics, and a median value for RMSE of all models for `by trial' are presented in Table \ref{tab:trial-avg-table}. The values are computed per participant, where each participant is identified by a two-letter code in front of each model and had multiple (3 to 12) trials. The numerical values for RMSE show that biLSTM model performs the best of all four models on average. Additionally, it is the best-performing model for most of the trials (69 out of 108 trials), when the results for each trial are observed separately. The worst-performing model, on average, is the CNN+RNN. 

We present below three trials that give us insight into the performance of the `by trial' experimental setup.

\begin{table}[th]
    \centering
    \includegraphics[width=0.55\linewidth]{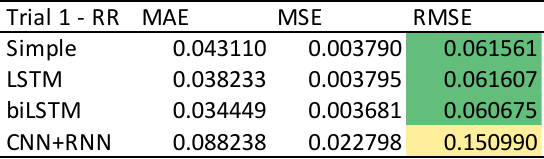}
    \caption{Setup by Trial: Trial 1 - RR Error Metrics}
    \label{tab:trial-1-table}
\end{table}

\begin{figure}[th]
    \centering
    \includegraphics[width=0.99\linewidth]{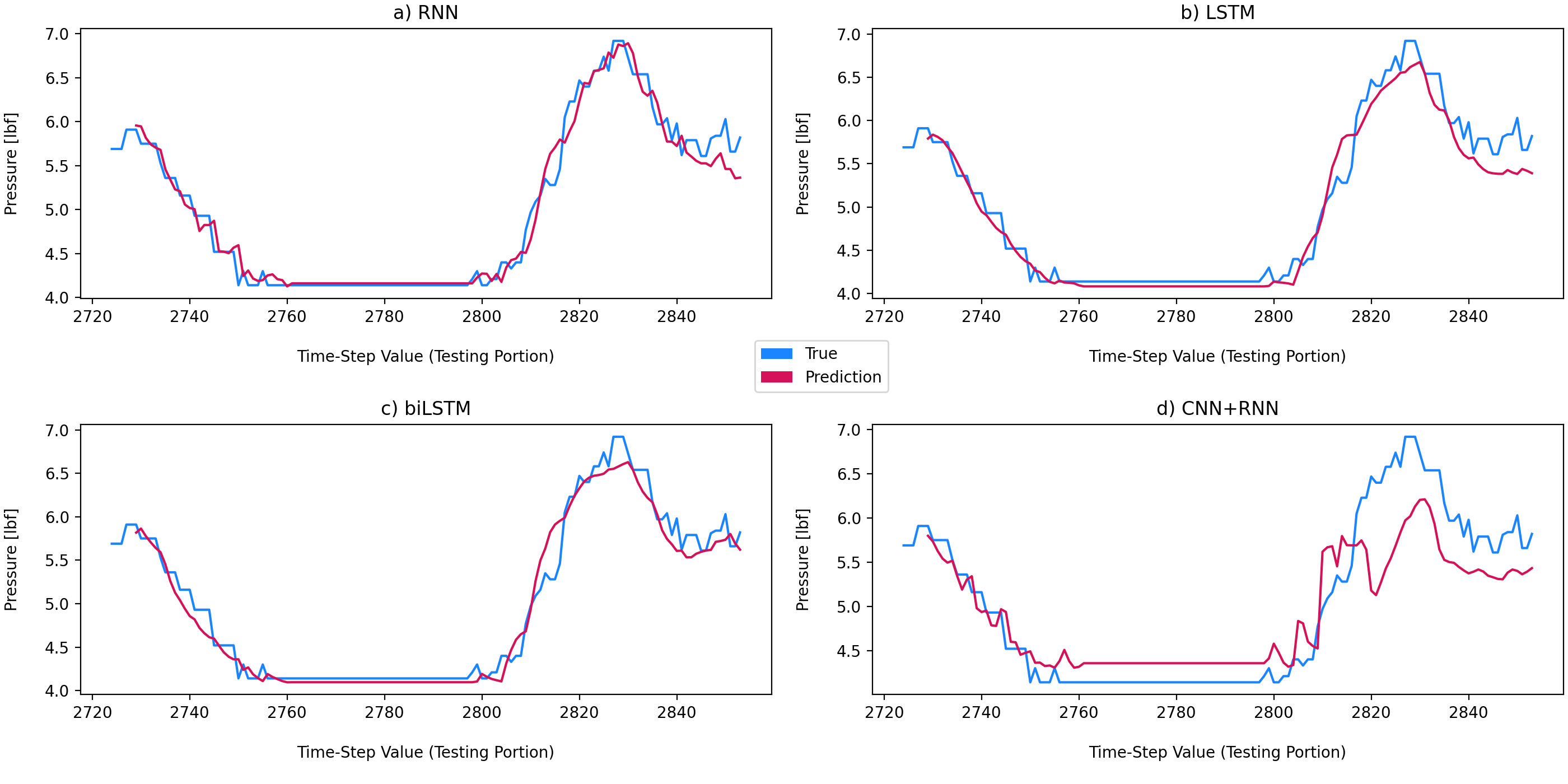}
    \caption{Setup by Trial: Trial 1 - RR Summed up Values for all (Sensors) True Test Data Values (blue) vs Predicted Data Values (red)}
    \label{fig:trial-1-sum}
\end{figure}

\paragraph{Trial 1 by Participant RR}
The results for this trial follow a similar trend to the experiment's overall average, where the biLSTM models perform the best, followed by RNN, LSTM, and lastly, CNN+RNN. Table \ref{tab:trial-1-table} shows the numerical error values for a trial from participant RR (referred to as Trial 1 - RR). 

This trial, at face value, is the best performing amongst all the experiments in setup `by trial,` with the lowest RMSE score of 0.060675 for the biLSTM model. Other RNN models show comparable performance, while the CNN+RNN has a noticeable struggle with prediction accuracy, showing a much higher RMSE value than the other three models. The plots of true test data values and predicted values (Figures \ref{fig:trial-1-sum} a) - d)) clearly show this observation. Figures \ref{fig:trial-1-sum} a) - c) visualize the better performance of the RNN models (RNN, LSTM, and biLSTM) compared to the CNN+RNN model shown in Figure \ref{fig:trial-1-sum} d). The CNN+RNN shows a much larger difference between the true test data values and the prediction data values compared to the RNN models (RNN, LSTM, and biLSTM). Visual inspection of raw data values (plot not included here) shows that two of the sensors experience significantly smaller variations in the testing portion of the truncated data than in the training/validation portion of the data. Hence, the pattern during testing is much different than the established pattern during the training and validation, which prevents CNN+RNN from more accurate predictions.  

\paragraph{Trial 2 by Participant AJ}
\begin{table}[th]
    \centering
    \includegraphics[width=0.55\linewidth]{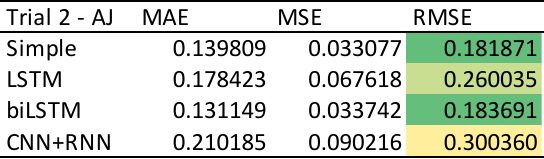}
    \caption{Setup by Trial: Trial 2 - AJ Error Metrics}
    \label{tab:trial-2-table}
\end{table}

\begin{figure}[th]
    \centering
    \includegraphics[width=0.99\linewidth]{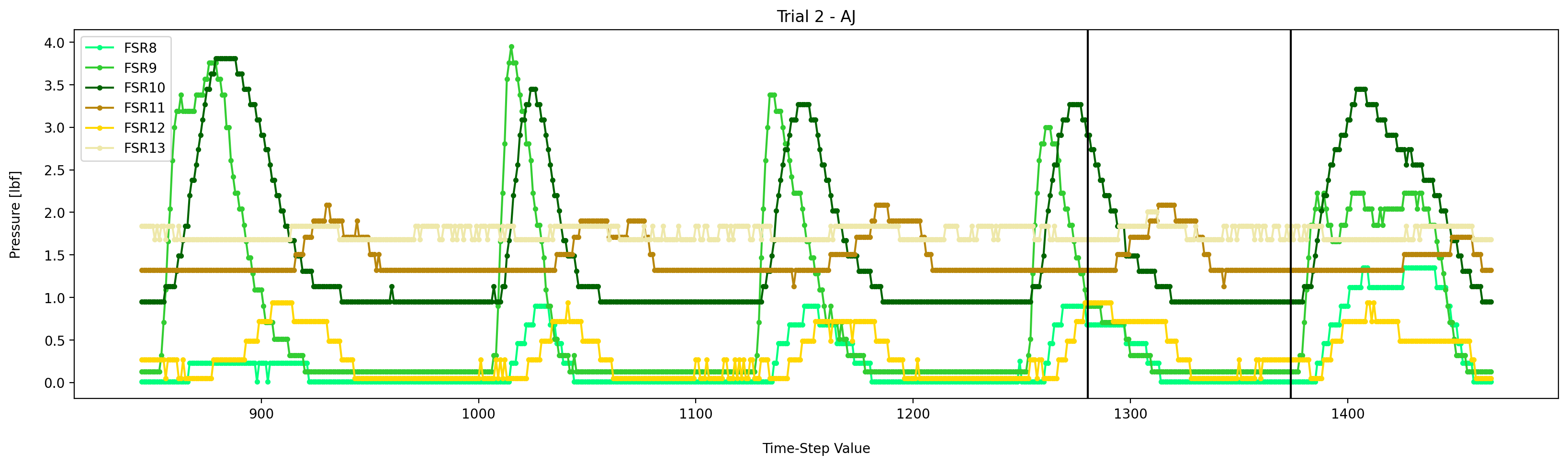}
    \caption{Setup by Trial: Trial 2 - AJ}
    \label{fig:trial-2}
\end{figure}

\begin{figure}[th]
    \centering
    \includegraphics[width=0.99\linewidth]{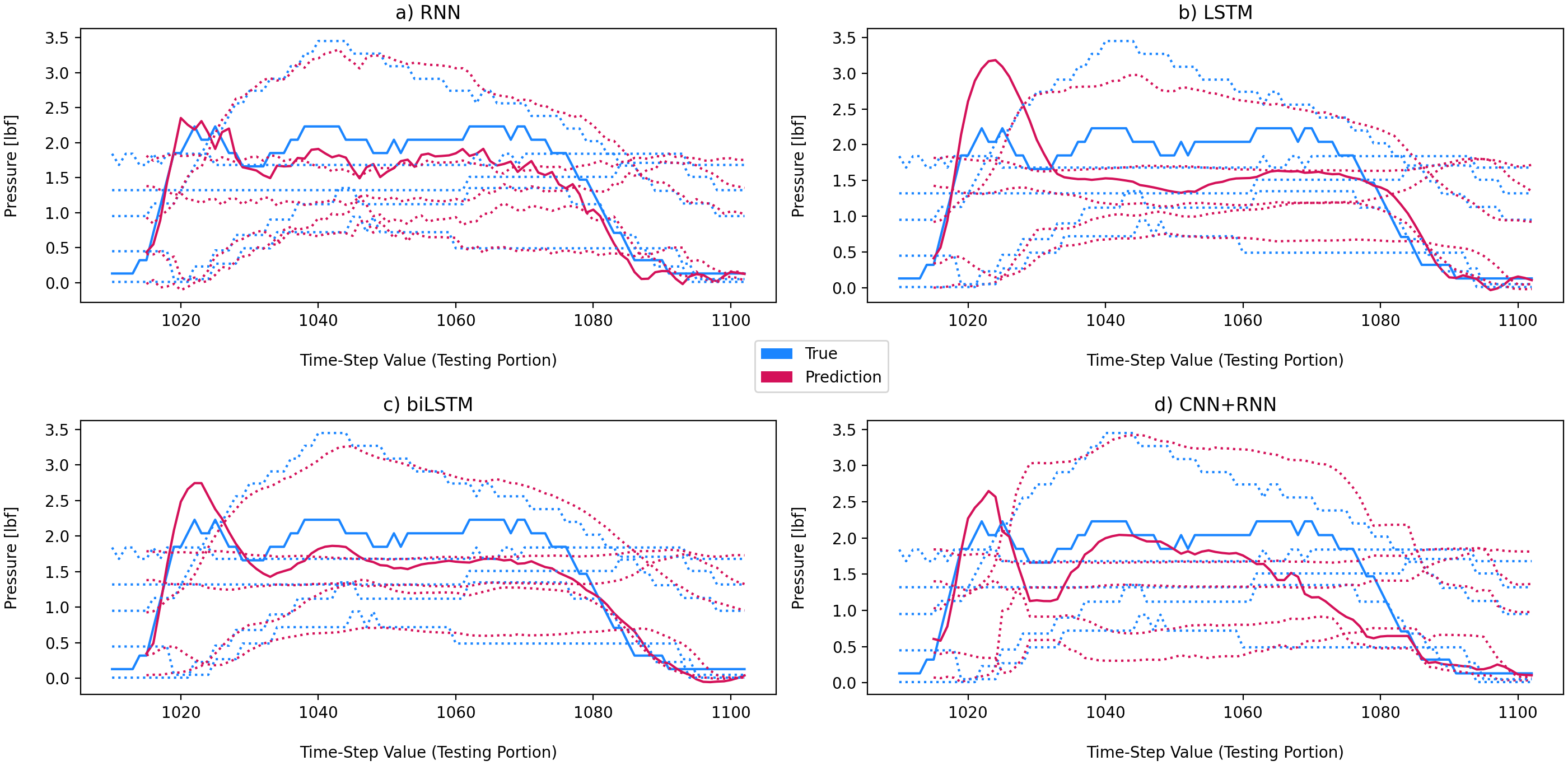}
    \caption{Setup by Trial: Trial 2 - AJ Individual Values for all (Sensors) True Test Data Values (blue) vs Predicted Data Values (red)}
    \label{fig:trial-2-individual}
\end{figure}

Table \ref{tab:trial-2-table} shows the numerical error values for one trial from participant AJ (referred to as Trial 2 - AJ). This trial tests the adaptability of all the models to predict data that greatly differs from the training input. To explore the behavior of all models under these circumstances, we look into the truncated data for the individual sensors shown in Figure \ref{fig:trial-2}. The fluctuation of the values in the training or validation datasets for FSR 9 are vastly different from the fluctuations in the testing portion. Moreover, for FSR 8, FRS 10, and FSR 11 the values for the testing dataset also slightly differ from the training and validation sets.  Figure \ref{fig:trial-2-individual}. shows separately plotted true test data values (blue) and predicted values (pink) for each sensor, where the values for FSR9 have been highlighted using solid lines. 
The predicted values for all models do not follow closely the true test data values. The difference in patterns of training and validation data compared to true test data values (Figure \ref{fig:trial-2}) gives insight into poor performance. This trial underlines the setbacks for the CNN+RNN and the advantages of the RNN models (specifically the RNN and biLSTM models). The RNN and biLSTM models perform adequately at predicting the true test data values. The separately plotted data in Figure \ref{fig:trial-2-individual} shows that both models struggle at correctly predicting FSR 9 but to some extent capture the overall trend of the rest of the sensor data. The CNN+RNN, on the other hand, struggles with predicting the values for most of the sensors.

\paragraph{Trial 3 by Participant AJ}

\begin{table}[th]
    \centering
    \includegraphics[width=0.55\linewidth]{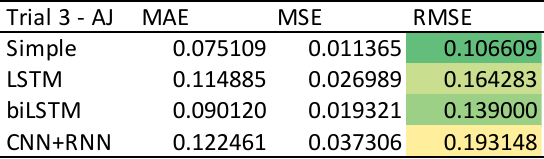}
    \caption{Setup by Trial: Trial 3 - AJ Error Metrics}
    \label{tab:trial-3-table}
\end{table}

\begin{figure}[th]
     \centering
     \includegraphics[width=0.99\linewidth]{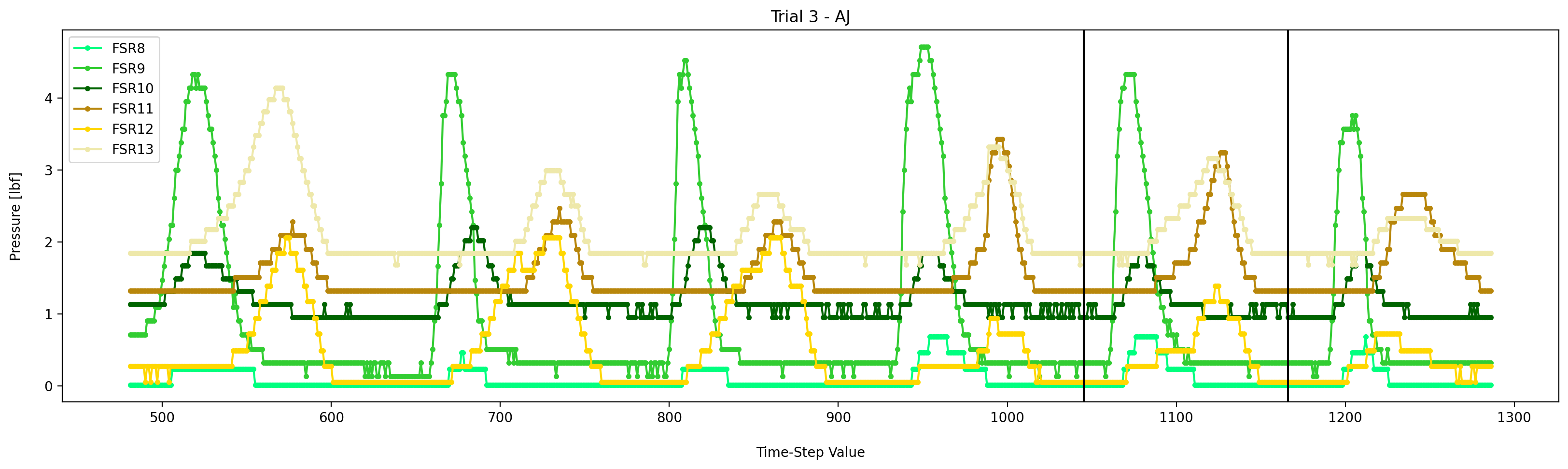}
     \caption{Setup by Trial: Trial 3 - AJ}
     \label{fig:trial-3}
 \end{figure}

Table \ref{tab:trial-3-table} shows the numerical error values for the last trial presented in this experimental setup, another (different) trial from participant AJ (referred to as Trial 3 - AJ). In this trial, the simple RNN outperforms all models, and the CNN+RNN (expectantly) performs poorly. As seen in Figure \ref{fig:trial-3} the last (testing) segment of data has a significantly different pattern than the training data. In this trial, FSR 9 and FSR 11 have vastly different values during testing from the corresponding values in training. We need to be cognizant of the applications of our method. The models will likely have much better performance if the true test data is allocated from a segment that corresponds to the established rhythm of walking, and not from the end/slowdown, where the walking pattern differs.

This observation points to another direction for future experimental work: we might be able to use the value of RMSE as an indicator of change of pattern for the participant within a single trial. We would train the models with the data collected at the beginning of walking, hence, we would capture the established pattern. As a participant walks for a longer time, they might experience fatigue, and therefore a change of pattern. The change of pattern could be detected and its quantitative measurement (RMSE) could be reported. This application would be suitable for evaluation and recovery monitoring.

\textit{Discussion:} The trials in our dataset are comprised of varying recorded lengths of time when an individual walks. After truncating the data to ensure the initial and concluding readings were removed (please see Section \ref{sec:prepossessing}), the length of data for each trial varied even further. The longer the trial, the more data was allocated for each portion: training, validation, and testing. As a result of having more data in the testing dataset, there is a chance for the testing dataset to be comprised of a swing phase along with a stance phase. During the swing phase of the gait cycle, the sensors are not in contact with the ground, resulting in relatively flat (small) sensor values. These relatively flat values are easy for the models to predict, which will favorably influence the RMSE and other error metrics. With smaller trials, and, thus, smaller testing portions, there is less chance of a swing phase to be included in the testing portion, which may lead to that model performing much worse in comparison to the longer trials. Trial 1 - RR is one such example where the testing portion includes a swing phase. 
Additionally, we trained the model on the data that corresponds to the beginning, validated the data towards the end, and tested it at the very end of the trial. The implications are that the training and validation were done when the participant had established a rhythm during walking. However, the testing is done as they are slowing to a stop. Therefore, the pattern of walking corresponding to the testing data may differ from the pattern for the data corresponding to training and validation, sometimes having fewer changes in data values. Using Figure \ref{fig:trial-1-sum} d) we can observe that the CNN+RNN model does not perform well during the stance phases, and the swing phase of the prediction is likely helping to reduce the error metrics. With this in mind, it is important to note that the RMSE and other error metric scores, while indicative of accuracy for a specific trial, should not be the only means to compare prediction accuracy from different trials. A swing phase, which is a natural part of the gait cycle, may be included in the test portion of one but not in the test portion of another trial, thus resulting in a smaller RMSE for the trial whose true test data values include the swing phase. The accuracy of prediction for the `by trial' setup will depend on the inclusion of the swing or stance phase in testing data.

We can make a few general observations: the CNN+RNN model may be less capable at short-distance prediction when the pattern in testing is slightly different than the pattern learned from training, and all models have an `advantage' at predicting when the data in the true test dataset has fewer fluctuations (swing phase). Nevertheless, the RNN, LSTM, and biLSTM models show very good accuracy for short-distance prediction for the data where the values of the training dataset vary in a slightly different way than in testing (Figure \ref{fig:trial-1-sum} a) - c)). This observation motivates our future work, where we would like to explore detecting pattern changes for non-healthy participants while walking. One idea is developing a deep-learning model to detect changes from the established pattern in real-time, by means of increasing errors between the predicted and newly generated data (corresponding to the testing portion). To peruse this line of work we need to evaluate the prediction run-time (please see Section~\ref{sec:limitations_and_future}) and carefully consider short- vs. long-distance prediction capabilities. 

Overall, the short-distance prediction using RNN models shows promising results and warns us of the importance of data capturing during experimentation in the (physical) laboratory environment and the selection of training, validation, and testing datasets during modeling and testing.

\subsubsection{Setup by Participant}
\label{sec:exp_by_participant }

\begin{table}[th]
    \centering
    \includegraphics[width=1.0\linewidth]{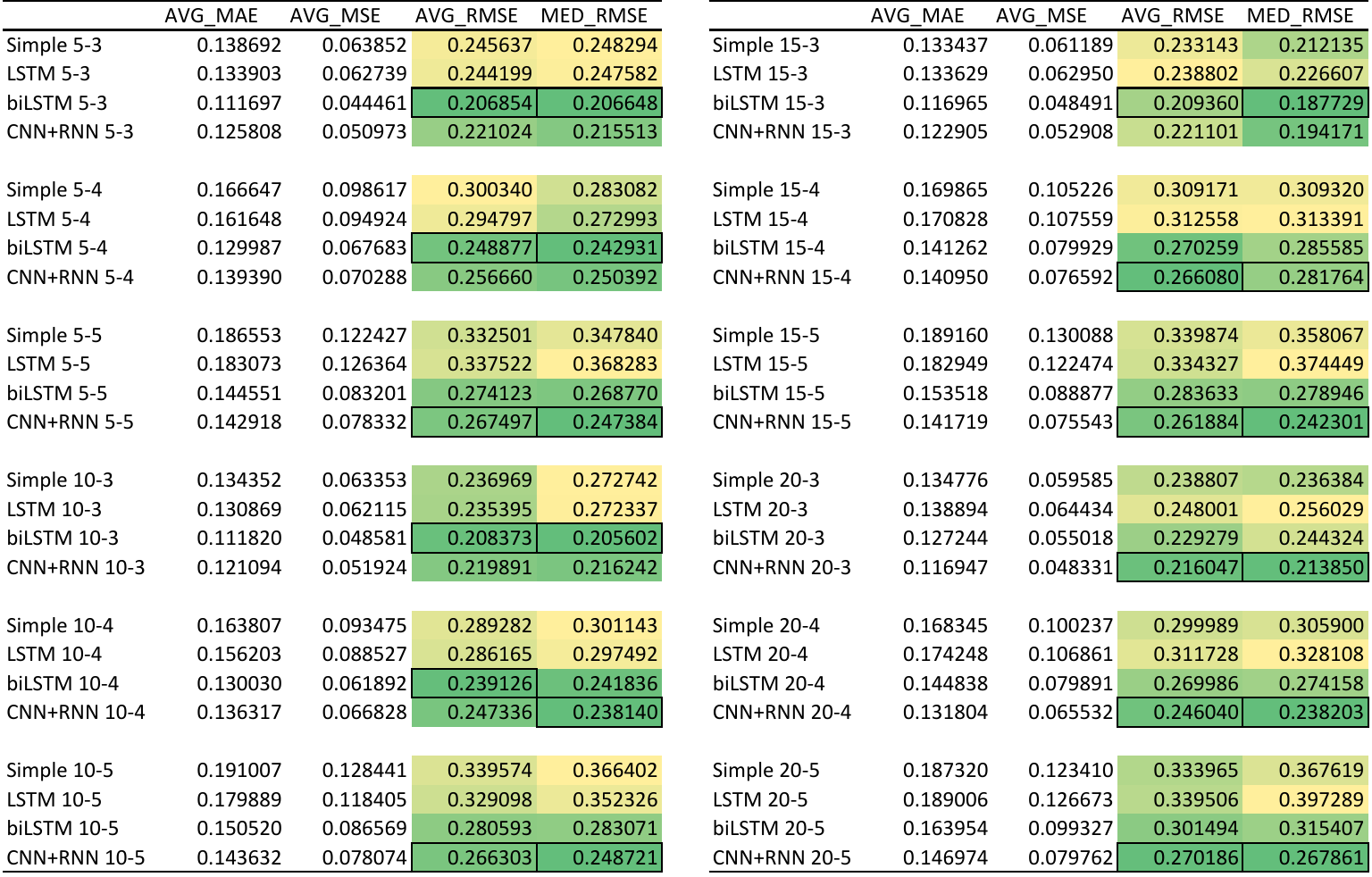}
    \caption{Setup by Participant: Average Error Metrics  by Model}
    \label{tab:participant-avg}
\end{table}

The average values for all 3 metrics for each experiment in the `by participant' setup while varying window size for both input and output (prediction) are shown in Table \ref{tab:participant-avg}. The input window size takes a value from a set (5, 10, 15, 20) and an output window from a set (3, 4, 5) for each input size, thus forming a total of 12 combinations for each participant and for each model. We also calculate and present the median value for RMSE as this experiment compares results for prediction for different participants. The tables follow the same shading scheme (dark green - best, pale yellow - poorest), and in addition, the outline shows the best, and lowest values for average and median values for RMSE, respectively. 

The numerical results indicate that for smaller input window sizes, 5-3, 5-4, 10-3, 10-4, and 15-3, the biLSTM model performs the best. However, for all other, larger, input window sizes the CNN+RNN models perform the best. These results align with the intuition behind the models. The biLSTM uses all input data to formulate the prediction, and therefore it performs very well for small window sizes. In the CNN+RNN model, the convolutional layers decrease the size of the input data by passing the data through filters while keeping the information pertinent to the prediction. Conversely, with larger input windows, the convolutional layers can extract the pattern much faster and easier while discarding the excess data, hence showing better performance. The RNN models do not have this capability. With the larger window sizes the biLSTM and the other RNN models, struggle with the increased input data size, while the CNN+RNN excels. We present the results for a total of six participants:  the first four (Tables \ref{tab:participant-1-error} - \ref{tab:participant-3-error}) present the best result for each of the considered input sizes, where the output size is selected based on the best RMSE score for CNN+RNN model. The next participant is selected to showcase an outlier: a case where biLSTM outperforms CNN+RNN by a small margin (Table \ref{tab:participant-4-error}). The last participant included a set of results (Tables \ref{tab:participant-5-error}) illustrating how much better the CNN+RNN model performs for larger inputs compared to the RNN models.

\begin{table}[th]
    \centering
    \includegraphics[width=0.60\linewidth]{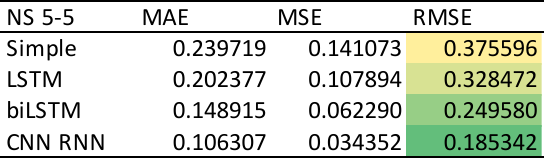}
    \caption{Setup by Participant: NS 5-5 Error Metrics}
    \label{tab:participant-1-error}
\end{table}

\begin{figure}[th]
    \centering
    \includegraphics[width=0.99\linewidth]{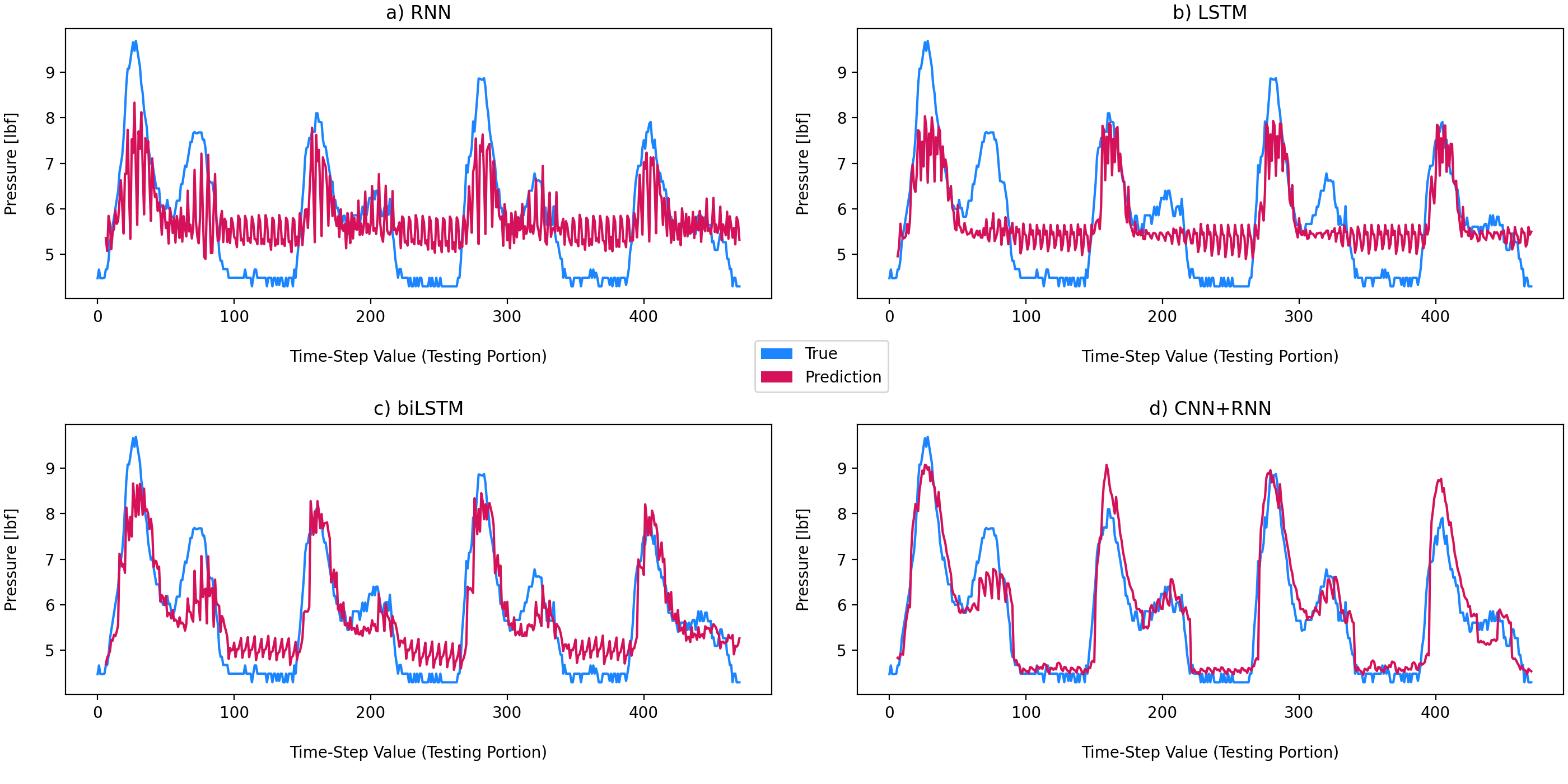}
    \caption{Setup by Participant: NS 5-5 Summed up Values for all (Sensors) True Test Data Values (blue) vs Predicted Data Values (red)}
    \label{fig:participant-1-sum}
\end{figure}

\paragraph{Participant NS 5-5}
Table \ref{tab:participant-1-error} shows the numerical error values for participant NS for input window value of 5 and output window value of 5 (referred to as NS 5-5), the participant with best performing CNN+RNN 5-5 model. For a short input, the intuition would be that the CNN+RNN would suffer due to a lack of input data. However, based on the average values, the CNN+RNN model performs best for this window for the large majority of participants for 5-5 experiments. All three RNN models underperform compared to CNN+RNN. We argue that while the input window is considerably small (i.e. manageable for RNN models), the large output window (5) in combination with the small input window makes it difficult for the RNN models to accurately predict. The sum of true test data values (sensor values) and predicted data values are shown in Figures \ref{fig:participant-1-sum}.a) - \ref{fig:participant-1-sum}.d) for the four observed models for NS 5-5. The cause for the large values for RMSE for the RNN and LSTM models is clearly visible in Figures \ref{fig:participant-1-sum} a) and b) as a miss-match between true test and predicted data values. The prediction with the smaller values of RMSE shows a much better fit as seen in Figures \ref{fig:participant-1-sum}.c) and \ref{fig:participant-1-sum}.d), with CNN+RNN (Figures \ref{fig:participant-1-sum}.d)) being the most accurate.

\begin{table}[h]
    \centering
    \includegraphics[width=0.60\linewidth]{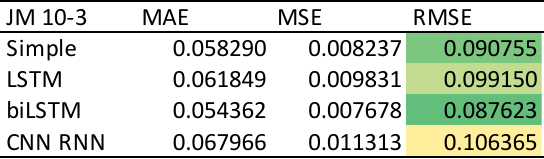}
    \caption{Setup by Participant: JM 10-3 Error Metrics}
    \label{tab:participant-2-error}
\end{table}

\begin{figure}[h]
    \centering
    \includegraphics[width=0.99\linewidth]{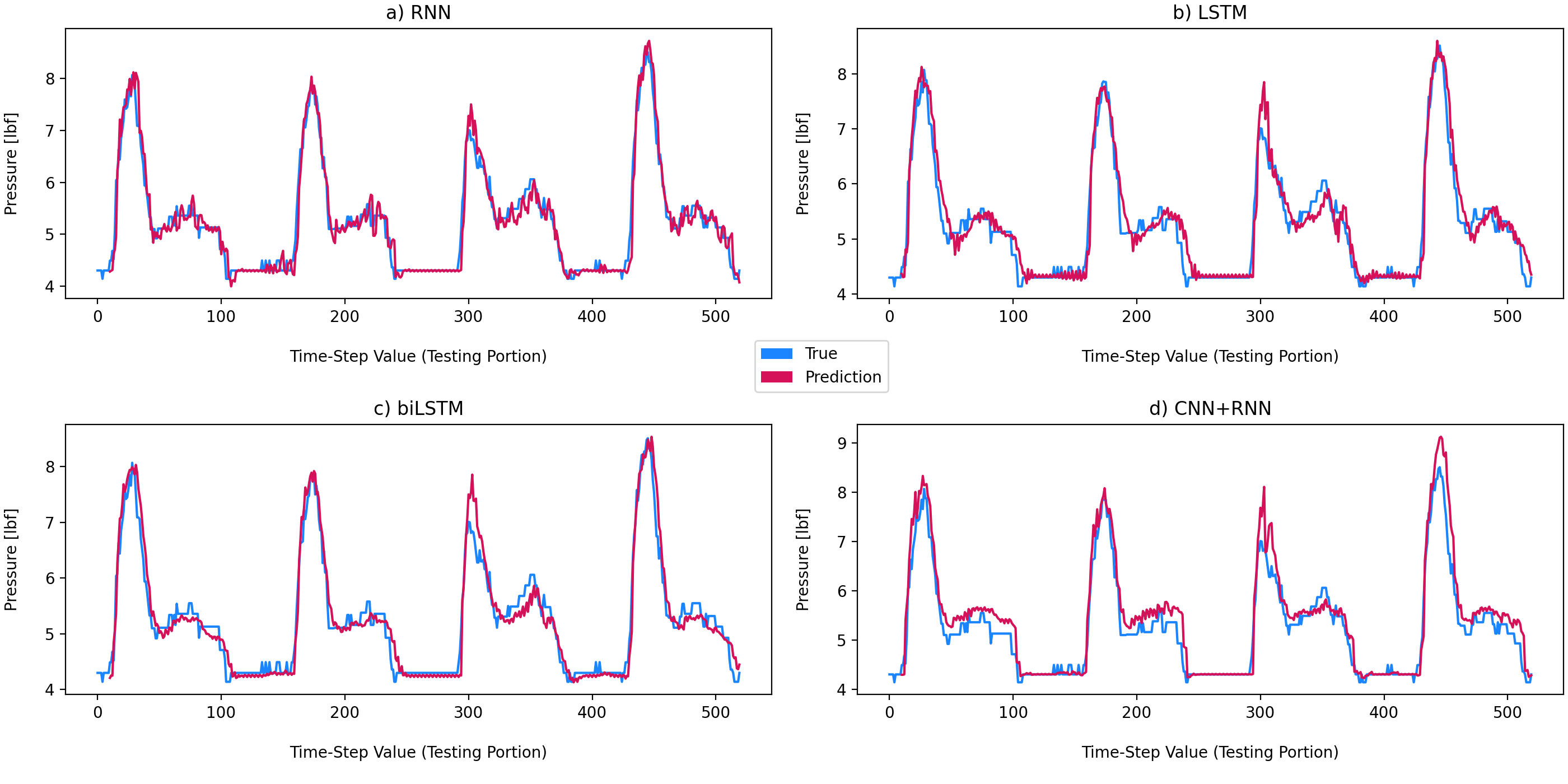}
    \caption{Setup by Participant: JM 10-3 Summed up Values for all (Sensors) True Test Data Values (blue) vs Predicted Data Values (red)}
    \label{fig:participant-2-sum}
\end{figure}

\paragraph{Participant JM 10-3}
Table \ref{tab:participant-2-error} shows the numerical error values for participant JM for an input window value of 10 and output window value of 3 (referred to as JM 10-3). This set of results is from the best-performing participant JM (best overall performance for all participants and all input/output window sizes) and shows that each of the models performs similarly well, with the CNN+RNN model performing slightly worse than the RNN models (RMSE for CNN+RNN model is by 0.018742 greater than RMSE for the best-performing model biLSTM). CNN+RNN does not exhibit its usual best performance. For this participant, the CNN+RNN  performance is slightly worse than other RNN models, as the output size is fairly small (3 rather than 5) for CNN+RNN to be efficient. Figures \ref{fig:participant-2-sum}.a) - \ref{fig:participant-2-sum}.d) show that the prediction closely follows the true test data values for all models. biLSTM is the best-performing model for this participant in input/output window size (Figure \ref{fig:participant-2-sum}.c).

In general, all four models for JM 10-3 perform better than previous models for participant NS (NS 5-5).  Specifically, the CNN+RNN model for JM 10-3 performs better than all models for participant NS 5 - 5 due to increased input window size: from 5 (for NS 5-5) to 10 (for JM 10-3). 

\begin{table}[h]
    \centering
    \includegraphics[width=0.6\linewidth]{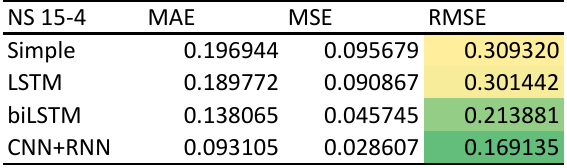}
    \caption{Setup by Participant: NS 15-4 Error Metrics }
    \label{tab:participant-3-1-error}
\end{table}

\begin{figure}[h]
    \centering
    \includegraphics[width=0.99\linewidth]{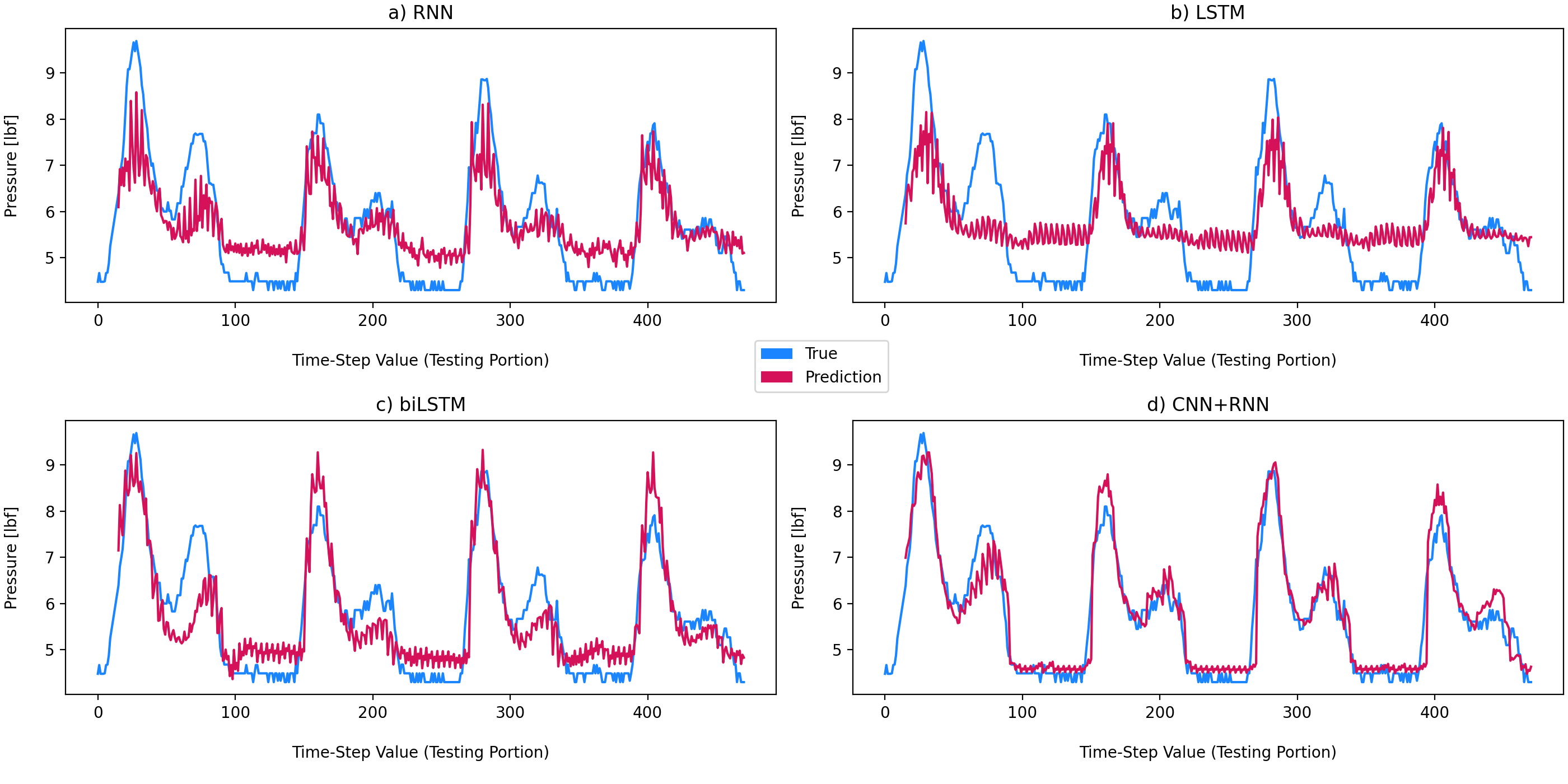}
    \caption{Setup by Participant: NS 15-4 Summed up Values for all (Sensors) True Test Data Values (blue) vs Predicted Data Values (red)}
    \label{fig:participant-3-1-sum}
\end{figure}

\paragraph{Participant NS 15-4} Table \ref{tab:participant-3-1-error} shows the numerical error values for participant NS for an input window value of 15 and an output window value of 4 (referred to as NS 15-4). This is the participant that has the best-performing model for the input size of 15. The smallest RMSE is obtained for the CNN+RNN model, and this model significantly outperforms all other RNN models. The input window size is large, making it difficult for RNN models to generate accurate predictions. Although CNN+RNN is the best-performing model, its RMSE is slightly larger than for JM 10-3. We hypothesize that the input size of 15 is not sufficient for accurate prediction for 4 consecutive values. Figures \ref{fig:participant-3-1-sum}.a) - \ref{fig:participant-3-1-sum}.d) visualize the prediction values for all 4 models. We can observe a better fit for CNN+RNN prediction compared to RNN predictions. Additionally, the visualization confirms the slightly poorer performance of all models compared to the models for JM 10-3.

\begin{table}[h]
    \centering
    \includegraphics[width=0.6\linewidth]{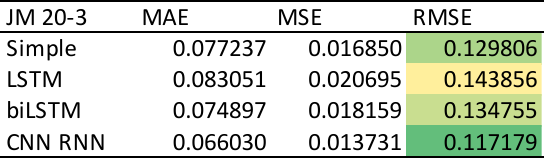}
    \caption{Setup by Participant: JM 20-3 Error Metrics}
    \label{tab:participant-3-error}
\end{table}

\begin{figure}[h]
    \centering
    \includegraphics[width=0.99\linewidth]{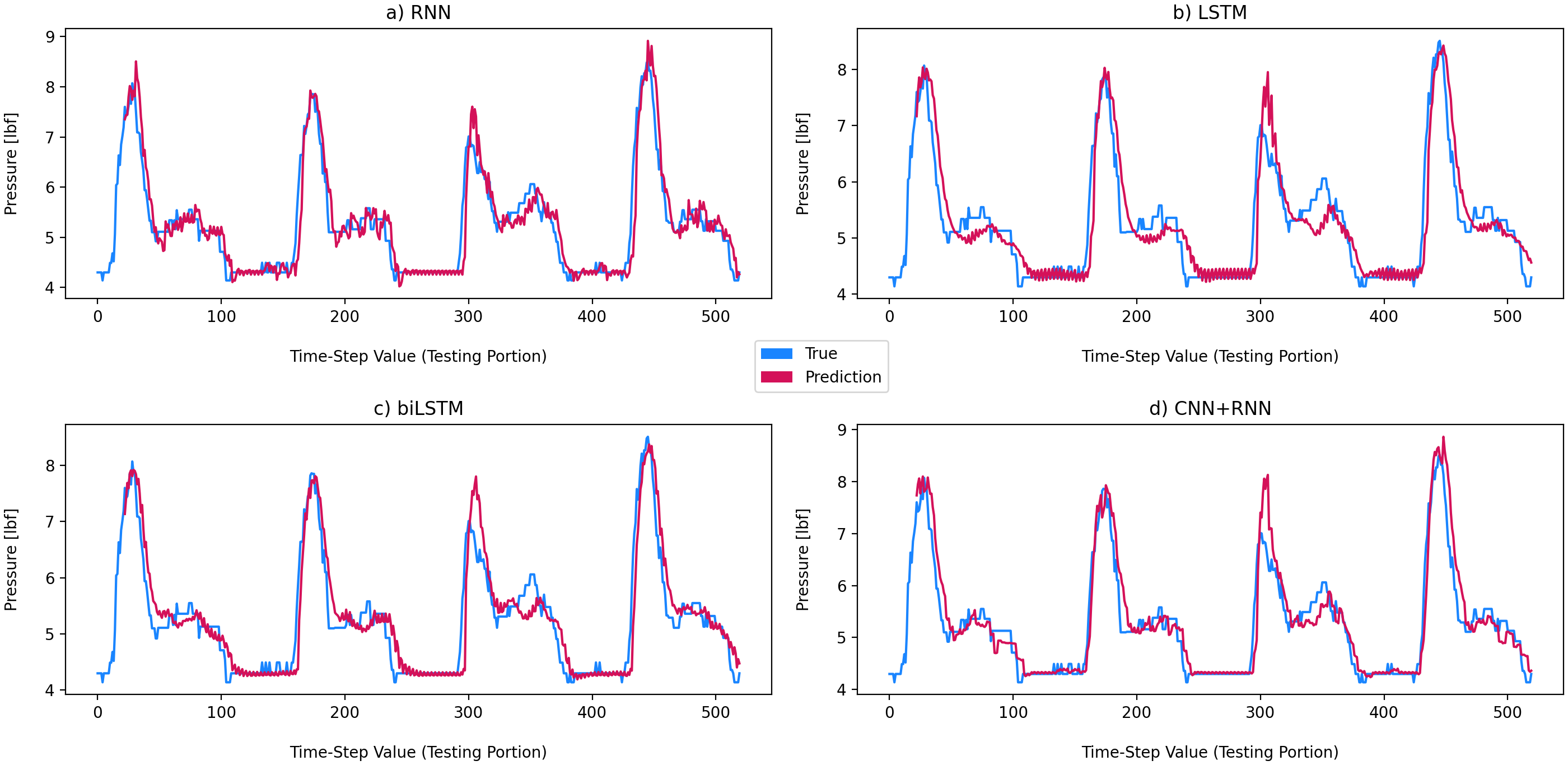}
    \caption{Setup by Participant: JM 20-3 Summed up Values for all (Sensors) True Test Data Values (blue) vs Predicted Data Values (red)}
    \label{fig:participant-3-sum}
\end{figure}

\paragraph{Participant JM 20-3} Table \ref{tab:participant-3-error} shows the numerical error values for participant JM for an input window value of 20 and output window value of 3 (referred to as JM 20-3). Because the input window is so large, the CNN+RNN does a great job at finding patterns in the data, leading it to perform the best even with the small output window. At the same time, the input size becomes too large for RNN models to generate accurate predictions. The RNN models have slightly higher values of RMSE compared to the values for JM 10-3. Figures \ref{fig:participant-3-sum}.a) - \ref{fig:participant-3-sum}.d) visualize the prediction values for all 4 models. They show a noticeably better fit for CNN+RNN prediction compared to RNN predictions.

\begin{table}[h]
    \centering
    \includegraphics[width=0.6\linewidth]{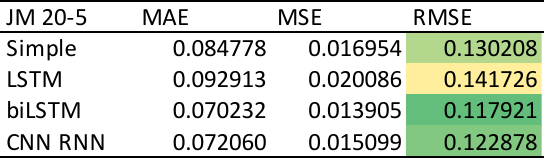}
    \caption{Setup by Participant: JM 20-5 Error Metrics}
    \label{tab:participant-4-error}
\end{table}

\paragraph{Participant JM 20-5 } Table \ref{tab:participant-4-error} shows the numerical error values for participant JM for an input window value of 20 and output window value of 5 (referred to as JM 20-5). We present participants JM because they have the best RMSE value for CNN+RNN for a window size of 20-5 compared to other participants (for the same window size). Surprisingly, the biLSTM model slightly outperforms the CNN+RNN model (by about 0.005). The large input and large output window should allow the CNN+RNN model to perform the best, as it does for every other participant for this input and output window combination. With such a small difference in RMSE, it might still pay off to use CNN+RNN as it takes a shorter time to train, validate, and test than the biLSTM model (please see Table \ref{tab:participant_runtimes}).

\begin{table}[th]
    \centering
    \includegraphics[width=0.609\linewidth]{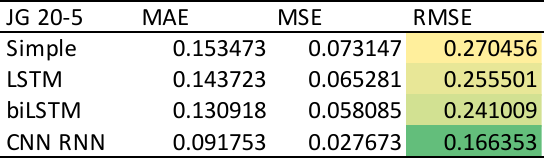}
    \caption{Setup by Participant: JG 20-5 Error Metrics}
    \label{tab:participant-5-error}
\end{table}

\paragraph{Participant JG 20-5} 
Table \ref{tab:participant-5-error} shows the numerical error values for participant JG for an input window value of 20 and an output window value of 5 (referred to as JG 20-5). These results are the closest to the average values for all four models, as presented in Table \ref{tab:participant-avg}. The CNN+RNN model outperforms all RNN models: simple RNN, LSTM, and biLSTM. CNN+RNN model excels due to the large input window size of 20 that provides additional history and context that helps the generation of accurate prediction for the large output size of 5. The large input size is an obstacle for RNN models that cannot keep up the accuracy for predicting as much as 5 consecutive output values. 

\textit{Discussion:} Our results indicate that the CNN+RNN models may be more capable and more suitable at \textit{predicting long sequences of time-series data} than traditional advanced RNN methods. All of our experiments show that an increasing number of trials used for training (i.e. increased training length) in the `by participant' setup does not correlate with the models having higher accuracy. We observe that the accuracy depends on the characteristics of the individual's gait, which we leverage in our custom prediction models. Furthermore, all models for long-distance prediction are able to predict the entire trial in less than 0.11 sec, while CNN+RNN models took  0.088 sec, on average. Not only that the CNN+RNN models generate prediction in a shorter time, but also in the vast majority of cases they generate predictions with lower errors. We observed that in a few cases, CNN+RNN models slightly underperform compared to the biLSTM models (for example in JM 20-5). With a very small difference in accuracy, the use of the CNN+RNN model may be a good tradeoff because it takes considerably less time to train and predict.

Additionally, as per our results, our models are able to \textit{predict entire trial}. Our models are trained on varied numbers of concatenated trials, where one trial is used for validation, and one is used for testing. Consequently, the models make predictions for the  \textit{entire trial} with high accuracy for the setup `by participant' designed for long-distance prediction (Table \ref{tab:participant-avg}). This observation shows the applicability of the proposed value prediction to the target applications: for long-term or in-home monitoring and prognosis of recovery, fall prediction, or to aid the exoskeleton movement. For those applications, the training and validation would be done in a clinical setting (after the participant's therapy or training sessions) when a desired goal has been achieved. Therefore, the models would capture Pareto-optimal gait patterns. The test data would be captured in an in-home setting. For recovery or fall prediction, the magnitude of the error between the predicted and captured data would indicate changes in the participant's health or abilities. For aiding the exoskeleton movement, the predicted data values could be used to continue the sequence of exoskeleton movements after the participant initiates the movement in a certain direction. Additionally, the `by participant' setup is suitable for authentication, where the training would be done for a set of known participants, and the errors between predicted and newly captured values would quantify the likelihood of the match to any of the known participants. The lack of match within a defined threshold can point to an unknown subject.

Our models are unique because they are customized and they predict values, rather than being trained for the entire participant population and classifying the participant's trials into predefined classes. This is not only a contribution but also an indication of the applicability of our models for the above-mentioned applications.

\section{Limitations and Future Directions}
\label{sec:limitations_and_future}

We propose future work to address the limitations of this study, but also to explore future software and hardware system design for a wider range of populations and applications.

Our study was performed using the data collected from healthy participants with an age range from 18 to 28, using data from the reference (right) foot.  We will expand our work to include a wide range of ages and different health conditions while exploring the differences in gait (pressure) patterns for both legs. We will explore how the insight gained from value prediction can be used for diagnosis, in addition to progress assessment, rehabilitation, aiding exoskeleton movement, or authentication for individuals with different health conditions.

Our run-time data are estimates as they include the system overhead of the Google Cloud Platform. They are computed for each experiment: where for each trial/participant the training and validation are run once, and testing is run 1,000 times and the results are averaged. We recognize that the used hardware provides substantial computation, communication, and storage resources, hence we might not be able to achieve this speed on a more limited platform.  Since the value prediction is targeted to be used in real-time applications, we need to extend the proposed approach to software system design where the cloud, edge, or micro-controller could host training or run the models to evaluate the performance. 
Due to considerations like data privacy, autonomy, and cost, Liberis et al. \cite{Liberis_22} contain machine learning inference within the device itself. However, deep learning, especially training, with limited hardware resources is an extremely challenging problem. We plan to evaluate the trade-offs for communication, memory utilization, performance, battery life, and wearability for the future system.

Additionally, we will explore the application of different filters for raw data, such as a simple running average filter, which could improve training time, resource requirements, and testing run time. Burdack et al. \cite{Burdack2020} used filtering, time derivatives, and data scaling to preprocess the data obtained from Ground Reaction Force sensors. Such preprocessing was used for classification, but we may explore the use of some of those methods for value prediction. Additionally, deep learning itself could be used for feature extraction and data preprocessing \cite{Morbidoni2019, Costilla-Reyes2021}.

While running our deep learning models, we did not employ N-fold cross-validation. We are questioning its application for real-time scenarios, as it implies random selection of a portion of a trial (for `by trial' setup) or one of the trials (for `by participant' setup). This kind of selection is not feasible for real-time applications as it needs prediction generated for the most recent trial, rather than a random trial. N-fold cross-validation is a possible approach for non-real-time applications, like authentication.

Our stretch goal is to develop an easy-to-use, inexpensive sensor-based system, accompanied by appropriate deep learning models to accurately assess movement disorders in gait, accurately predict fall risks, or aid exoskeleton movement for a wide range of individuals. We will research the use of several different sensor types, placed on various body parts to improve prediction accuracy or extend applicability to other movement or postural disorders. Finally, we plan to implement a conditional Generative Adversarial Network and Attention mechanism to further preserve long-distance information loss. We also intend to make use of the pre-trained state-of-the-art deep neural network models to fine-tune our time series gait data.

\section{Conclusions}
\label{sec:conclusion}
We present a technique for the application of \textit{customized} deep-learning models for a value prediction of gait spatiotemporal parameters, namely pressure sensor data. We compare and contrast short- and long-distance predictions in two different experimental setups, `by trial' and `by participant', respectively. As our models expand beyond classifiers, so do their applications. We target their use in a variety of health-related applications as well as for biometric authentication. The novelty of our work also extends to the application of deep learning models to time series of gait data, which despite the smaller number of features have high prediction accuracy.

Our results show that the best-performing model for a short sequence prediction is Bidirectional LSTM with an average  RMSE = 0.124346 and a minimum RMSE = 0.060675. The best-performing model for long-distance prediction in the vast majority of cases is CNN+RNN with an average RMSE = 0.194171, and as low as RMSE = 0.106365. Moreover, the proposed CNN+RNN setup can accurately predict an entire trial, when trained and validated using the trials from the same participant. We illustrate prediction accuracy by plotting the sum of sensor values for each time step and the sum of predicted data for the corresponding time step.

Our future directions will be in the domains of exploring and possibly including a variety of sensors, fine-tuning the acquisition technique, expanding the approach to a wider population, and designing corresponding deep learning models to prove accuracy for extended application domains. Our goal is to explore hardware and software system design to achieve ease of use, accuracy, and affordability that will use gait data for health, well-being, and security applications.

\section*{Acknowledgement}
We would like to acknowledge Grayson Hill and Jessica Wei for their help with the engineering setup and discussion, and John Sanchez for his early work on surveying the related literature.

We would also like to thank all participants who volunteered for this study.

\bibliographystyle{unsrtnat}
\bibliography{references} 

\end{document}